\documentclass[pdflatex,sn-mathphys-num]{sn-jnl}
\usepackage{graphicx}%
\usepackage{multirow}%
\usepackage{amsmath,amssymb,amsfonts}%
\usepackage{amsthm}%
\usepackage{mathrsfs}%
\usepackage[title]{appendix}%
\usepackage{xcolor}%
\usepackage{textcomp}%
\usepackage{manyfoot}%
\usepackage{booktabs}%
\usepackage{algorithm}%
\usepackage{algorithmicx}%
\usepackage{algpseudocode}%
\usepackage{listings}%
\usepackage{float}
\usepackage{soul}
\usepackage{upgreek}
\usepackage{epstopdf}
\def\bk{{\boldsymbol{k}}}
\def\bq{{\boldsymbol{q}}}
\def\bG{{\boldsymbol{G}}}
\def\br{{\boldsymbol{r}}}
\def\H{\mathcal{H}}

\unnumbered

\begin{document}

\title[Article Title]{Superfluid stiffness of twisted multilayer \\ graphene superconductors}

\author[1]{\fnm{Abhishek} \sur{Banerjee}}
\equalcont{These authors contributed equally to this work.}

\author[1]{\fnm{Zeyu} \sur{Hao}}
\equalcont{These authors contributed equally to this work.}

\author[1]{\fnm{Mary} \sur{Kreidel}}
\equalcont{These authors contributed equally to this work.}

\author[1]{\fnm{Patrick} \sur{Ledwith}}

\author[1]{\fnm{Isabelle} \sur{Phinney}}

\author[2]{\fnm{Jeong Min} \sur{Park}}

\author[1]{\fnm{Andrew} \sur{Zimmerman}}

\author[3]{\fnm{ Kenji} \sur{ Watanabe}}

\author[4]{\fnm{Takashi} \sur{Taniguchi}}

\author[1]{\fnm{Robert} \sur{M Westervelt}}

\author[2]{\fnm{Pablo} \sur{Jarillo-Herrero}}

\author[1,5]{\fnm{Pavel} A. \sur{Volkov}}

\author[1]{\fnm{Ashvin} \sur{Vishwanath}}

\author*[6]{\fnm{Kin} \sur{Chung Fong}}\email{kc.fong@rtx.com}

\author*[1]{\fnm{Philip} \sur{Kim}}\email{pkim@physics.harvard.edu}

\affil[1]{\orgdiv{Department of Physics}, \orgname{Harvard University}, \orgaddress{\city{Cambridge}, \postcode{02138}, \state{MA}, \country{USA}}}

\affil[2]{\orgdiv{Department of Physics}, \orgname{Massachusetts Institute of Technology}, \orgaddress{\city{Cambridge}, \postcode{02138}, \state{MA}, \country{USA}}}

\affil[3]{\orgdiv{Research Center for Electronic and Optical Materials}, \orgname{National Institute for Materials Science}, \orgaddress{\street{1-1 Namiki}, \city{Tsukuba}, \postcode{305-0044}, \country{Japan}}}

\affil[4]{\orgdiv{Research Center for Materials Nanoarchitectonics}, \orgname{National Institute for Materials Science}, \orgaddress{\street{1-1 Namiki}, \city{Tsukuba}, \postcode{305-0044}, \country{Japan}}}

\affil[5]{\orgdiv{Department of Physics}, \orgname{University of Connecticut}, \orgaddress{\state{MA}, \postcode{06269}, \country{USA}}}

\affil[6]{\orgdiv{Raytheon BBN Technologies}, \orgaddress{\state{MA}, \country{USA}}}

\abstract{
The robustness of the macroscopic quantum nature of a superconductor can be characterized by the superfluid stiffness, $\rho_s$, a quantity that describes the energy required to vary the phase 
of the macroscopic quantum wave function. 
In unconventional superconductors, such as cuprates, the low-temperature behavior of $\rho_s$
drastically differs from that of conventional superconductors due to quasiparticle excitations
from gapless points (nodes) in momentum space. Intensive research on the recently discovered magic-angle twisted graphene family has revealed, in addition to superconducting states, strongly correlated electronic states associated with spontaneously broken symmetries, inviting the study of $\rho_s$ to uncover the potentially unconventional nature of its superconductivity. Here we report the measurement of $\rho_s$ in magic-angle twisted trilayer graphene (TTG), revealing unconventional nodal-gap superconductivity. Utilizing radio-frequency reflectometry techniques to measure the kinetic inductive response of superconducting TTG coupled to a microwave resonator, we find a linear temperature dependence of $\rho_s$ at low temperatures and nonlinear Meissner effects in the current bias dependence, both indicating nodal structures in the superconducting order parameter. Furthermore, the doping dependence shows a linear correlation between the zero temperature $\rho_s$ and the superconducting transition temperature \(T_c\), reminiscent of Uemura's relation in cuprates, suggesting phase-coherence-limited superconductivity. Our results provide strong evidence for nodal superconductivity in TTG and put strong constraints on the mechanisms of these graphene-based superconductors.
}

\maketitle

Superconductivity arises from the condensation of pairs of electrons, known as Cooper pairs, and is characterized by a complex order parameter, comprising an amplitude and a well-defined phase. Microscopically, the amplitude of the order parameter, $\Delta$, is related to the strength of Cooper pairing, whereas the superfluid stiffness, $\rho_s$, characterizes the phase-rigidity of the condensate order parameter. Measurements of $\rho_s$ are highly motivated not only because phase rigidity is a defining property of the superconducting state, responsible for both zero resistance and the Meissner effect, but also because they help distinguish between conventional and unconventional superconducting orders~\cite{lee1997unusual,emery1995importance}. The dependence of $\rho_s$ on temperature~\cite{lee1997unusual, MILLIS19981742} and superfluid velocity ~\cite{yip1992nonlinear,xu1995nonlinear} carries distinctive signatures of the superconducting pairing symmetry. For instance, the linear temperature dependence of \(\rho_s(T)\) in cuprates is the hallmark of nodes in the superconducting order parameter, i.e., vanishing $\Delta$ along a specific direction in momentum space, resulting from the unconventional pairing in d-wave superconductivity in cuprates~\cite{hardy1993precision}. The magnitude and doping dependence of $\rho_s$ also offer deep insights into the details of the superconducting gap and the underlying electron bands \cite{lee1997unusual}. Notably, in a large class of unconventional superconductors, $\rho_s$ is orders of magnitude smaller than in conventional Bardeen-Cooper-Schrieffer (BCS)-type superconductors and follows unusual scaling laws~\cite{uemura1989universal} relating superconducting phase fluctuations with \(T_c\)~\cite{emery1995importance,keimer2015quantum}. 

The recent discovery of magic-angle twisted multilayer graphene (MATMG) ~\cite{cao2018unconventional,hao2021electric,park2021tunable,park2022robust,zhang2022promotion} has provided an unprecedented opportunity to study the emergence of superconductivity due to strongly correlated electrons. However, the intense investigations thus far have relied primarily on electrical transport and scanning-tunneling-spectroscopy (STS) measurements, and the superconducting phase remains poorly understood. Electrical transport experiments varying displacement field, magnetic field, and carrier density have shown nematicity~\cite{cao2021nematicity}, Pauli limit violation~\cite{cao2021pauli}, unusual behavior of the superconducting critical current~\cite{tian2023evidence},  and correlation with symmetry breaking in the superconducting state~\cite{cao2018unconventional,hao2021electric,park2021tunable}, suggesting a paradigm beyond BCS physics. However, zero resistance of the superconducting state precludes any exploration of the internal structure of the superconducting dome using transport. On the other hand, STS studies showed V-shaped differential conductance spectra, consistent with nodal superconductivity~\cite{oh2021evidence,kim2022evidence}. Yet this single-particle tunneling interpretation is complicated by a difficulty in distinguishing superconducting gaps from gaps due to non-superconducting orders such as flavor polarized correlated states. The need to better understand the superconducting order parameter in MATMG calls for measurements of thermodynamic quantities in the superconducting states such as \(\rho_s\). In particular, measuring \(\rho_s\) allows one to probe inside the superconducting dome, reveal the doping evolution of the superconducting state, and provide insights into the origin of superconductivity. However, this has been technically challenging due to the reduced sample dimensionality and size. Only indirect estimates of \(\rho_s\) based on critical-current density analyses are available so far~\cite{tian2023evidence}.

In this work, we use radio-frequency (rf) resonant circuits~\cite{schoelkopf1998radio, reilly2007fast,vigneau2023probing,crossno2016observation}
to measure the superfluid response of micrometer-sized TTG superconductors. The microwave response of a superconductor at finite temperatures can be captured by a two-fluid model~\cite{tinkham2004introduction} — a parallel circuit of two channels: the superfluid channel composed of condensed Cooper pairs, and the normal fluid channel composed of quasi-particle excitations. The frequency ($f$) dependent complex rf conductivity in this model can be described by a dissipative real part, $\sigma_1=n_n e^2\tau_n/m_e$ and dissipationless imaginary part, $\sigma_2=n_s e^2/2\pi fm_e$, where $e$, $m_e$ are electron charge and mass, respectively, $n_e$ and $n_s$ are the normal and superfluid electron densities and $\tau_n$ is the scattering time for quasi-particle excitations. 
For the frequency regime $f<f_c=(n_s/n_n)/2\pi\tau_n$, $\sigma_1\ll \sigma_2$, we expect the impedance of the superconducting sample to be dominated by a dissipationless inductive channel with the characteristic inductance \(L_K = 1/(2\pi f \sigma_2)\), often referred to as the kinetic inductance. \(L_K\) is related to the superfluid stiffness by $\rho_s = (w/l)(\hbar^2/4e^2L_K)$, where \(w\) and \(l\) are the device width (perpendicular to the current flow) and length, respectively. Therefore, by incorporating the superconducting TTG sample into a microwave resonator with a resonance frequency that reflects \(L_K\), we can measure its superfluid stiffness.

\begin{figure}[h]
\centering
\includegraphics{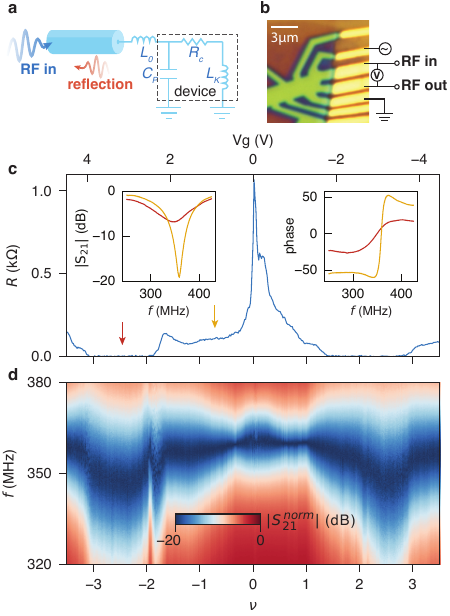}
\caption{{\bf Experimental setup and device characterization:} \textbf{a,} Schematic of the radio-frequency (rf) reflectometry setup. An LC matching network consisting of $L_0$ and $C_P$ transforms the device impedance $1-10$k$\Omega$ to the 50~$\Omega$ characteristic impedance of the RF measurement circuit. \textbf{b,} Optical microscope image of the twisted-trilayer graphene device reported in the main text and a simple schematic of the measurement scheme. \textbf{c,} DC resistance $R$ of the device as a function of moiré filling factor $\nu$. The sample is superconducting ($R=0$) in both the electron and hole-doped sectors. Insets show the amplitude and phase response of the complex reflection co-efficient $S_{21}$ showing shifts in resonance frequency between the normal and superconducting states. \textbf{d,} 2D map of the normalized amplitude $|S^{\text{norm}}_{21}|$ shown as a function of frequency $f$ and filling factor $\nu$. For clearer visualization, for each \(\nu\), we define its $|S^{\text{norm}}_{21}(f)|= 1-(\mathrm{max}(|S_{21}(f)|)-|S_{21}(f)|)/0.9$. The resonator undergoes large frequency shifts $\Delta f_r \simeq 10-15$~MHz when the sample changes from normal to superconducting states.}\label{fig1}
\end{figure}

Although techniques utilizing rf resonantors have been used to measure kinetic inductances~\cite{annunziata2010tunable, singh2018competition,phan2022detecting}, the large inherent contact resistance \(R_c\sim1-10~\mathrm{k\Omega}\) and parasitic capacitive coupling $C_P$ in 2D material devices (see Fig.~\ref{fig1}a) can completely damp out the resonance, thus forbidding the application of these techniques in measuring \(L_K\) of 2D material superconductors. We solve this problem by using an impedance-matching network and adding a surface-mount inductor \(L_0\). Under our experimental conditions, \(C_P\sim1-10~\mathrm{pF}\) and a choice of \(L_0\sim100-200~\mathrm{nH}\) allows the combined impedance of the resonant circuit, $Z_D\simeq L_0/(C_P R_C)$, to match the characteristic rf matching impedance \(Z_0=50~\Omega\), resulting in lumped-element microwave resonant circuits with a resonance frequency \(f_{r0} = 1/\sqrt{2\pi fL_0C_P} \sim 100-300~\mathrm{MHz}\). Changes of kinetic inductance produce relative shifts of the microwave resonance frequency $\Delta f_r/f_{r0}$, and they are related by a linear relationship $\Delta L_K = 2 R_c^2 C_P \left(\Delta f_r/f_{r0}\right)$ for expected values of $L_K \simeq 10-100$s of nH $\leq 0.1~R_c^2 C_P \simeq 1~\mathrm{\upmu}$H. (see Methods). We contrast our approach with other experimental methods to measure superfluid stiffness that do not readily integrate with 2D material flakes~\cite{annunziata2010tunable, singh2018competition}, have high operating frequencies~\cite{phan2022detecting}, or require macroscopic-sized samples~\cite{bovzovic2016dependence}.

Our TTG device consists of three layers of graphene stacked with alternating twist angles of $\pm\theta=1.55^\circ$~\cite{khalaf2019magic}, encapsulated within a layer of hBN and graphite on both sides, assembled on an insulating silicon substrate to minimize parasitic capacitance. The graphite layers serve as gate electrodes and can be independently controlled to tune the carrier density and displacement electric field \(D\). Fig.~\ref{fig1}b shows an optical microscope image of the device, along with the measurement scheme, where we perform simultaneous DC electrical transport and microwave measurements (see Methods). Fig.~\ref{fig1}c shows the 4-terminal DC resistance \(R\) as a function of filling factor \(\nu\), defined as the number of carriers per moiré unit cell. We observe superconductivity indicated by zero resistance in the range of filling factors $-3<\nu<-2$ for hole doping and $1.5<\nu<3$ for electron doping, with a 2D \(\nu-D\) phase diagram (Fig.~S1b) that is similar to our previous studies~\cite{hao2021electric,park2021tunable}. Simultaneously, we measure the complex reflection coefficient of the resonator $S_{21}$, defined as the ratio between the output (reflected) and input microwave voltage signal \(V_{out}/V_{in}\) (see Methods). The amplitude \(\left| S_{21}\right|\) as a function of \(\nu\) and frequency \(f\), as shown in Fig.~\ref{fig1}d, displays a dip around 350~MHz due to resonance through the whole range of \(\nu\), and the resonance shifts to a lower frequency by $\Delta f_r \sim 10-15$~MHz as the sample transits from normal to superconducting states. The insets of Fig.~\ref{fig1}c show two examples of the \(S_{21}\) amplitude and phase response at the normal and superconducting states, respectively, with the filling indicated by the arrows of the corresponding color. We use a circle fitting method~\cite{probst2015efficient} to accurately extract the resonance frequency $f_r$, from which we estimate $L_K$ and $\rho_s = (w/l)(\hbar^2/4e^2L_K)$. In this analysis, we employed $w/l \simeq 5$ based on the device geometry (see Methods).

\begin{figure*}[h]
\centering
\hspace*{-1.8cm}
\includegraphics{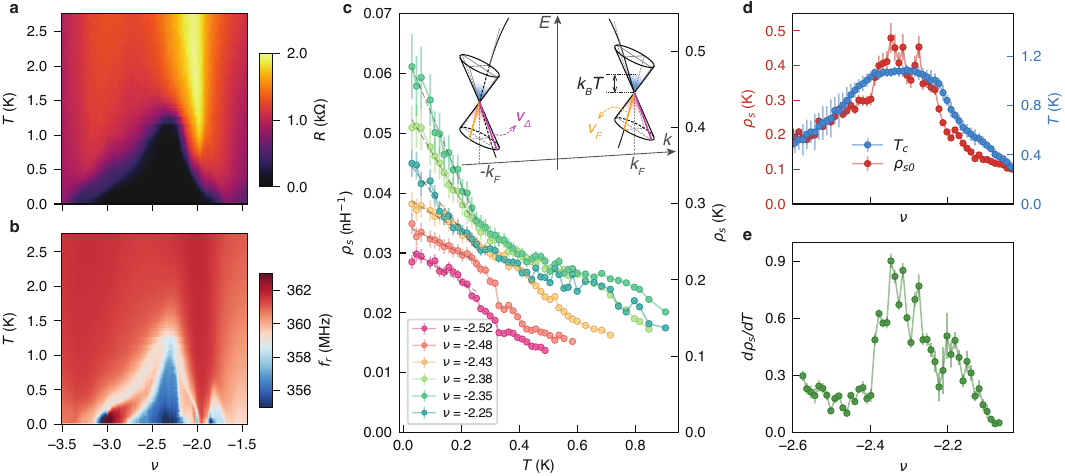}
\caption{{\bf Temperature and doping dependent superfluid stiffness:} Two-dimensional map of the \textbf{a,} DC resistance $R$ and \textbf{b,} resonator frequency $f_r$ measured as a function of moiré filling factor $\nu$ and sample temperature $T$. \textbf{c,} Superfluid stiffness $\rho_s$ as a function of \(T\) for a range of filling factors around $\nu=-2.4$. \textbf{d,} (Left-axis) Zero-temperature superfluid stiffness obtained by linear extrapolation of curves in \textbf{c,} as a function of $\nu$. (Right axis) $T_c$ obtained from DC resistance measurements as a function of $\nu$. \textbf{e,} Low-temperature slope $d\rho/dT$ obtained as a function of $\nu$.}\label{fig2}
\end{figure*}

Focusing on the hole-side superconductor, we measure \(R\) and \(f_r\) as a function of \(\nu\) and temperature \(T\). \(R(\nu,T)\)  shows a superconducting dome manifesting as the zero resistance area in the dark-colored region in Fig.~\ref{fig2}a, with a maximum \(T_c=1.2~\mathrm{K}\) at the optimal doping \(\nu_{op}=-2.3\), where $T_c$ is defined as the onset temperature for non-zero values of the DC resistance. Correspondingly, \(f_r(\nu, T)\) in Fig.~\ref{fig2}b also displays a dome of lower resonance frequency in about the same area. Converting \(f_r\) to \(\rho_s\), in Fig.~\ref{fig2}c we plot the temperature dependence of \(\rho_s\) in the units of both \(\mathrm{nH}^{-1}\) and Kelvin (K) at several filling factors around \(\nu_{op}\) within the superconducting dome. Across these filling factors, \(\rho_s(T)\) increases linearly with decreasing \(T\), with no sign of saturation down to the lowest measurable temperature \(\sim 30~\mathrm{mK}\), strongly indicating that the superconducting gap is nodal across the entire superconducting dome. We emphasize that this linear-in-\(T\) behavior occurs below $T\sim 0.2-0.5~\mathrm{K} \simeq 0.4T_c$, where the magnitude change of the order parameter is a subleading effect compared to the thermal activation of quasi-particles, and covers approximately an order of magnitude in temperature. At higher temperatures, \(\rho_s\) decreases in a reverse-S-shaped pattern and tends to flatten out, which coincides well with the onset of finite DC resistance (see Fig.~S3). 

The quasi-particle contribution to the linear-in-\(T\) suppression of $\rho_s(T) \propto n_s(T) \propto (n_s(0) - n_{qp}(T))$ implies a linear-in-\(T\) increase in the quasi-particle density, $n_{qp}$. This is not expected in a fully gapped superconductor, where the quasi-particle population is exponentially suppressed for  $T \leq 0.3~T_c$. Instead, it is consistent with a 2D nodal superconducting order, where the superconducting gap vanishes along certain directions — referred to as ``nodes" — on the Fermi surface~\cite{lee1997unusual, millis1998anomalous}, leading to a linear dispersion for the quasi-particle excitations. We illustrate this in Fig.~\ref{fig2}c inset, where we plot the superconducting quasi-particle energy dispersion near two nodes that are symmetrically located at the Fermi momentum \(\pm k_F\) in the lattice momentum \(k\)-space. Two distinct velocities $v_\Delta$, corresponding to the slope of the superconducting gap, and $v_F$, the Fermi velocity, describe the linear dispersion perpendicular and parallel to the axis connecting the two nodes, respectively. A non-zero \(T\) would immediately populate these cones with an equal number of quasi-particles and quasiholes, with their density linearly proportional to \(T\).

We further characterize the doping dependence of the low-temperature linear behavior of $\rho_s$ by performing a linear fit, as indicated by the dashed lines in Fig.~\ref{fig2}c. From this, we extract the zero-temperature limit of the superfluid stiffness $\rho_{s0}=\rho_s(T=0)$ and the slope $d\rho_{s}/dT$, shown in Fig~\ref{fig2}d and Fig~\ref{fig2}e, respectively. In Fig.~\ref{fig2}d, we compare \(\rho_{s0}\) with $T_c$ extracted from DC measurements. The two quantities roughly track each other in a bell-shaped curve centered around $\nu=-2.3$, suggesting that the same mechanism determines both $\rho_{s0}(\nu)$ and $T_c(\nu)$. On the other hand, the low-temperature slope $d\rho_{s}/dT$ is roughly constant $\sim0.2-0.3$ for \(-2.6<\nu<-2.4\), and then shows a sharp change rising to nearly $\sim 0.9$ around $\nu=-2.3$ before tapering off to $\sim 0$ as we approach the right edge of the dome at $\nu=-2$.

\begin{figure}[h]
\centering
\includegraphics{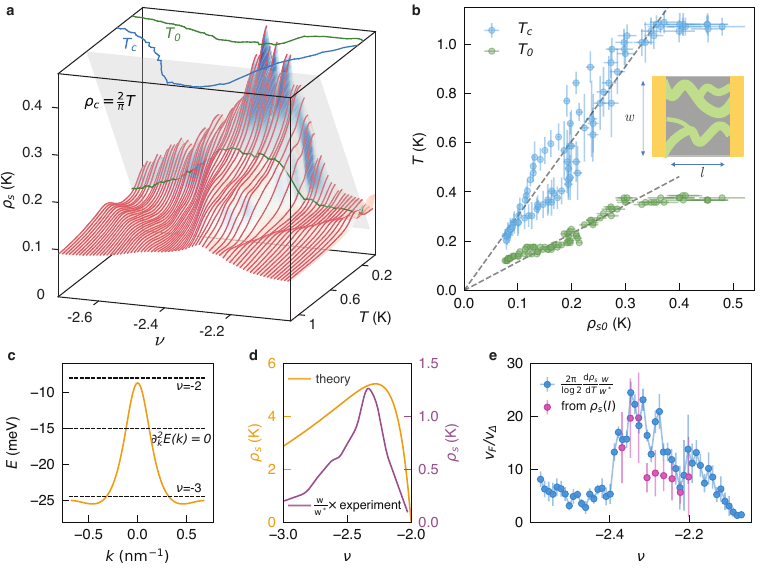}
\caption{{\bf BKT transition, Uemura's relation, and nodal pairing symmetry:} \textbf{a,} Waterfall plot showing $\rho_s$ as a function of $T$ and $\nu$. To aid visualization, the curves have been interpolated and smoothed with 6-pt moving average along the \(T\) axis. The BKT transition temperature ($T_{0}$) is obtained from the intersection between the universal BKT plane represented as $\rho_c=2T/\pi$ and the experimental stiffness curves. \textbf{b,} $T_c$ and $T_{0}$ plotted against $\rho_{s0}$. (Inset) Schematic showing inhomogeneous superconducting paths that may lead to an underestimation of the superfluid stiffness. \textbf{c,} Hartree-Fock renormalized bands in TTG as a function of momentum $k$ within the mini-Brillouin zone. The dashed lines mark the energy where \(\nu=-2, -3\) and the curvature \(\partial^2_kE(k)=0\) occur. \textbf{d,} Comparing the theoretical estimation of $\rho_{s0} (\nu)$ based on the Hartree-Forck renormalized band structure, with the experimentally obtained $\rho_{s0}(\nu)$, which is also interpolated and smoothed for clarity. Both theory and experiment show a bell-shaped $\nu$ dependence, with a lopsided maximum close to $\nu=-2$. Theory estimates are larger by $\times$5 compared to the experimental value. \textbf{e,} $v_F/v_\Delta$ as a function of $\nu$ obtained from the temperature dependence $\rho_s(T)$ and the current-bias dependence $\rho_s(I)$ (described later). The two methods roughly agree with $v_F/v_\Delta \simeq 2-20$.}\label{fig3}
\end{figure}

Considering that TTG superconductivity is a strictly 2D phenomenon, the experimentally obtained $\rho_s(T)$ allows us to estimate the Berezinskii–Kosterlitz–Thouless (BKT) transition temperature $T_0$ using the Nelson–Kosterlitz criterion: $T_{BKT}=\pi\rho_s(T_{BKT})/2$ ~\cite{nelson1977universal}. In the BKT theory, for $T>T_{BKT}$, the phase-coherence of a 2D superconductor is destroyed due to the proliferation of unbound vortices, restoring a finite resistance due to their dissipative motion. Experimentally, we can estimate the density-dependent BKT transition temperature from Fig.~\ref{fig3}a by locating the temperature $T_0$ where $\rho_s(T,\nu)$ (smoothed for clarity) intersects with the universal \(\rho_c=2T/\pi\) plane.  As shown in the top panel of Fig.~\ref{fig3}a, we find the $T_0(\nu)$ is smaller than $T_c$ obtained from the DC transport, indicating that the TTG device is superconducting even at $T>T_0$.
This observation is rather surprising as the simple BKT model seems to break down. While this observation could be explained by an unusually large vortex-core energy suppressing the proliferation of vortices $T>T_{BKT}$~\cite{bengatto2007kosterlitz,hetel2007quantum}, we find an explanation associated with sample inhomogeneity more plausible, as discussed below.

It is known that superconductivity in MATMG samples suffers from twist angle disorder ~\cite{turkel2022orderly}, which can create inhomogeneous superconductivity~\cite{hao2021electric}, and percolation paths when a supercurrent flows in the sample. Given such a network of superconducting paths, as illusrated in Fig.~\ref{fig3}b inset, the aspect ratio of the device $w/\ell$ for two-terminal DC and rf transport should be rescaled using the narrower effective sample width $w^*$. Therefore, our estimated $\rho_s$ from the measured kinetic response underestimates the true superfluid stiffness by a factor of $\alpha=w^*/w$. We can estimate \(\alpha\) from our experiment by noting the remarkable relationships between experimentally measured $\rho_{s0}$, $T_0$, and $T_c$, as shown in Fig.~\ref{fig3}b. In particular, we find $T_c/T_0\approx 3.0$ by taking the ratio between the slopes of the linear parts in $T_c$ versus $\rho_{s0}$ and in $T_0$ versus $\rho_{s0}$. Considering $T_{BKT}\approx T_c$ in MATMG~\cite{hao2021electric, park2021tunable} (i.e., zero resistivity for $T<T_{BKT}$), we estimate $\alpha=w^*/w\approx T_{BKT}/T_0 \approx T_c/T_0 \approx 3.0$

Having obtained $\alpha$, we can now make quantitative comparison between $\rho_{s0}$ and $T_c$. We find a linear correlation between $\rho_{s0}$ and $T_c$ away from the optimal doping — in both underdoped ($\nu>\nu_{op}$) and overdoped ($\nu<\nu_{op}$) regimes. A linear dependence of $T_c$ on $\rho_{s0}$ was first identified in underdoped cuprates ~\cite{uemura1989universal}, and subsequently observed in a variety of unconventional superconductors including overdoped cuprates~\cite{homes2004universal, dordevic2013organic, bovzovic2016dependence}, suggesting a universal scaling behavior known as the Uemura's relation. This phenomenological dependence suggests that long-range phase order, characterized by $\rho_{s0}$, determines $T_c$, in contrast to conventional BCS superconductors where the strength of Cooper pairing, characterized by the BCS gap $\Delta$, determines $T_c$. Following this interpretation and assuming parabolic band dispersions in TTG, we estimate  $T_c/E_F = T_c/(\alpha 4 \pi \rho_{s0}) \simeq 0.08$ based on the linear slope between $T_c$ and $\rho_{s0}$. This result indicates strongly-coupled superconductivity, consistent with more indirect analyses in previous works~\cite{hao2021electric,park2021tunable}. Note that for non-parabolic bands (see Supplementary Material), $E_F$ is larger and therefore the estimate will be somewhat lower. We have observed similar linear scaling relationships in other devices, in both electron and hole-doped sectors, even when effects of twist-angle disorder are evident (see Fig.~S7). This suggests the generality of this effect in MATMG superconductivity.

We now focus on the density-dependent behaviors of $\rho_{s0}$ and $d\rho_{s}/dT$ measured at the zero temperature limit. Recalling that $\rho_s=\hbar^2n_s/(4 m_e)$ for parabolic bands described by an effective mass $m_e$, we expect a linear relationship between $\rho_{s0}$ and $\nu \propto n_s$.
The bell-shaped dependence observed in the experiment [Fig.~2d], particularly the drop of $\rho_{s0}$ with increasing hole doping for $\nu<-2.4$, suggests a departure from the parabolic-band approximation. Theoretical modeling of the TTG band structure for moiré fillings between $-3 <\nu< -2$ suggests the presence of strongly renormalized band dispersion $E(k)$ that are highly non-parabolic, as shown in Fig.~\ref{fig3}c, obtained by Hartree-Fock calculation assuming flavor polarization of the TTG flat band. An interplay between strong Coulomb repulsion and quantum geometric effects, particularly the concentration of Berry curvature near $k=0$, leads to a band structure that is highly dispersive close to $\nu=-2$ but becomes progressively flatter on approaching $\nu=-3$. For such non-parabolic bands, a more general formula is required for the superfluid stiffness: $\rho_s (\nu)= \frac{1}{8} \sum_{{\bk}} n_k \nabla_{\bk}^2 E = \frac{v_F(\nu) k_F(\nu)}{8 \pi}$ (for an isotropic $E(\bk)$), leading to a doping dependence as shown in the theoretical estimate in Fig.~\ref{fig3}d, which is shown in yellow and plotted together with the smoothed experimental curve shown in red. Starting at $\nu=-2$, the initial steep rise of $\rho_{s}$ is enabled by a large positive band curvature $\nabla_{\bk}^2 E$ near $k=0$. However, on further doping,  $\nabla_{\bk}^2 E$ becomes negative, causing a reduction of the superfluid stiffness. Remarkably, the maximum of $\rho_s$ is reached well below the half-filling of the band due to its strong particle-hole asymmetry. However, the theoretical estimates are larger than the experimental values by a factor of $\times$5 even after the geometrical factor correction by $\alpha$. This discrepancy indicates that only a fraction of the electronic spectral weight is condensed into the superfluid. We leave the understanding of this to future studies. It is worth noting that a similar reduction of superfluid stiffness with increasing doping is observed in overdoped cuprates and remains a subject of active investigation~\cite{bovzovic2016dependence,mahmood2019locating}. 

With the geometrical correction $\alpha$, we can also now provide a quantitative analysis on the low temperature linear dependence of $\rho_s (T)$. Within the physical picture of nodal superconducting gaps, the aforementioned gap and electron band velocities \(v_\Delta\) and \(v_F\) are related to the linear slope of \(\rho_s(T)\) by $v_F/v_\Delta = (8\pi/N\log(2)) d\rho_s/dT \times \alpha$, where $N$ is the number of nodes (see Methods for details). Tentatively assuming $N=4$ (d-wave) and considering the scaling factor $\alpha=3$, we obtain $v_F/v_\Delta  \simeq 5-20$ as a function of $\nu$, as shown in Fig.~\ref{fig3}e. We also estimate $v_F/v_\Delta$ from current-bias dependent measurements  $\rho_s(I)$ (discussed below), and obtain similar values [Fig.~\ref{fig3}e], suggesting the robustness of our estimation of $v_F/v_\Delta$.

To solidify our interpretation of nodal superconductivity, we measure both \(R\) and $\rho_s$ in the presence of a finite DC supercurrent bias $I$. In nodal superconductors, a current \(I\) produces a Doppler shift of the quasi-particle spectrum due to finite superfluid velocity, resulting in a current-dependent quasi-particle population at the nodal points. This phenomenon is known as the non-linear Meissner effect and was initially proposed as a zero-temperature test for nodal superconductivity~\cite{yip1992nonlinear}, and observed in later experiments in cuprates~\cite{bidinosti1999magnetic,oates2004observation, wilcox2022observation}.

\begin{figure}[h]
\centering
\hspace*{-1.4cm}
\includegraphics{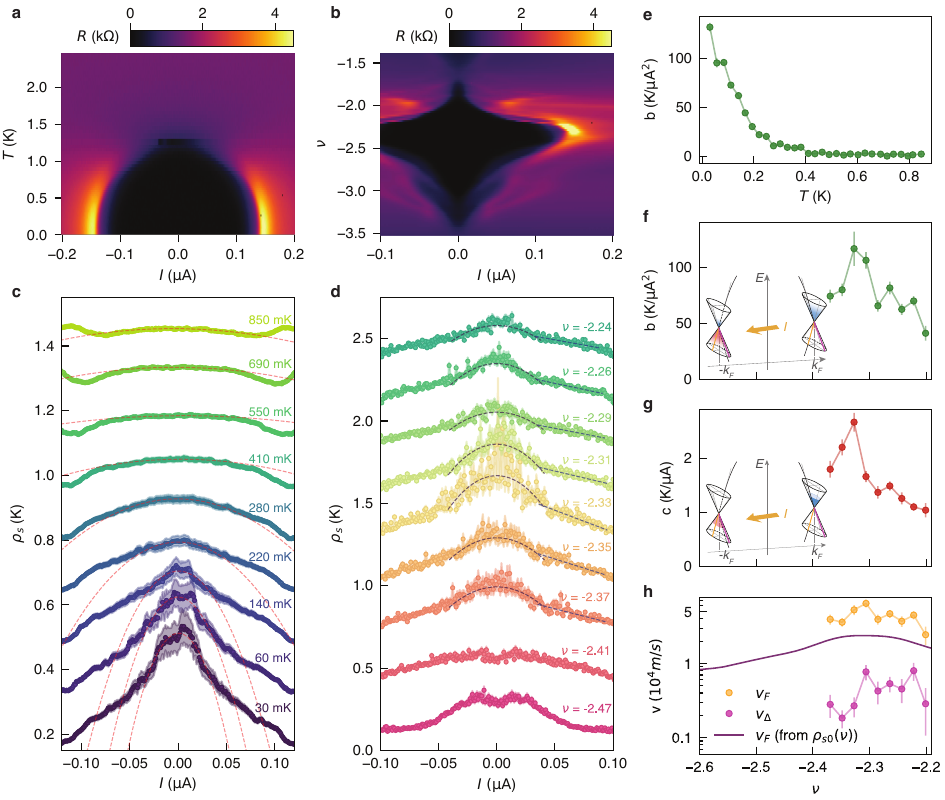}
\caption{{\bf Nonlinear Meissner effect:} \textbf{a,} Two-dimensional map of the DC resistance $R$ as a function of current bias $I$ and temperature $T$ close to optimal doping $\nu=-2.34$ where critical current, $I_c \simeq 0.12$~$\mathrm{\upmu}$A is maximum. \textbf{b,} $R$ as a function of supercurrent bias $I$ and moiré filling factor $\nu$. \textbf{c,} $\rho_s(I)$ measured at $\nu=-2.34$ at different values of sample temperature $T$. The curvature of $\rho_s(I)$ at zero current decays with increasing temperature. \textbf{d,} Superfluid stiffness $\rho_s$ as a function of $I$ for a range of filling factors $\nu=-2.24$ to $\nu=-2.47$ ($T=30$~mK). For a range of filling factors around optimal doping, $\rho_s(I) \propto -I^2$ for $I \to 0$ with a crossover to linear dependence $\rho_s(I) \propto -I$. The linear behavior persists for a large range of $I<I_c \simeq 0.12$~$\mathrm{\upmu}$A around $\nu=-2.34$. Curves are shifted for clarity. \textbf{e,} Temperature dependence of the curvature of $\rho_s(I)$ at $I=0$, $b(T)$. Diverging behavior is observed as $T \to 0$, a strong signature of nodal superconductivity. Base temperature value of \textbf{f,} the quadratic curvature $b$, and \textbf{g,} the linear slope, $c$, as a function of moiré filling factor. \textbf{h,} Fermi-velocity estimated from $\rho_s(I)$ and $\rho_s(T)$, indicating $v_F\simeq 0.1-0.7 \times 10^5$ m/s and $v_\Delta \simeq 0.2-0.5 \times 10^4$ m/s. Solid line shows $v_F$ calculated using the interpolated and smoothed experimental \(\rho_{s0}(\nu)\). The schematics in the insets of (f) and (g) depict the two different regimes of the nonlinear Meissner effect. In (f), \(I\to0\), both temperature and current produce quasiparticle excitations. In (g), \(I>I^*\), the current-induced effect dominates over temperature. In both cases, the current bias produces a finite population of quasi-particles in one node and quasi-holes in the opposite node. 
}\label{fig4}
\end{figure}

Figures~\ref{fig4}a and ~\ref{fig4}b show \(R\) as a function of \(I\) and \(T\), and a function of \(I\) and \(\nu\), respectively, with superconductivity indicated by the dark regions of zero resistance. Within the superconducting regime, we measure a rapid suppression of $\rho_s$ with increasing $I$ [Fig.~\ref{fig4}(c,d)], with the strongest signatures close to a range of filling factors around the optimal doping ($\nu=-2.3$, $I_c \simeq 0.12~\mathrm{\upmu}$A) [Fig.~\ref{fig4}d]. The suppression of $\rho_s$ follows a quadratic behavior, $\delta \rho_s=-bI^2$,  close to $I \to 0$,  which turns linear with $\delta \rho_s=-cI$ after \(I\) exceeds a cross-over scale $I^*$. We further investigate the temperature dependence of this effect at $\nu=-2.34$, close to the optimal doping, and find that the quadratic behavior at low bias is strongly temperature dependent [Fig.~\ref{fig4}c] with the quadratic coefficient $b$ appearing to diverge as $T \to 0$ [Fig.~\ref{fig4}e]. We also observe a non-trivial filling factor dependence of both the quadratic coefficient, $b(\nu)$, and the linear coefficient, $c(\nu)$, measured at $T=30$~mK [Fig.~\ref{fig4}(f,g)].

To understand these behaviors, we model the electrodynamic response of a 2D nodal superconductor in the presence of a DC supercurrent bias (see Methods), and obtain a quadratic dependence for $I\to0$:
\begin{equation}
\label{eqnl1}
\Delta \rho_s (I) = -\frac{v_F^3}{v_\Delta}\frac{1}{T}\frac{N}{128(4e\rho_{s0}w^*)^2} I^2
\end{equation}
and a linear dependence for $I>I^*$,
\begin{equation}
\label{eqnl2}
\Delta \rho_s (I) = -\frac{v_F^2}{v_\Delta} \frac{N \langle|\cos \theta_{I, v_n}|\rangle}{16(4e\rho_{s0}w^*)} I
\end{equation}
 with the crossover current $I^*(T)=(2e\rho_{s0}w^*/v_F)T$ separating the quadratic and linear regimes of $\rho_s(I)$. Here $\langle|\cos \theta_{I,v_n} |\rangle$ represents the angle between the applied supercurrent direction and the Fermi velocity at the node, averaged over all gap nodes.

 Our model captures all three experimental features: the quadratic suppression of stiffness at $I \to 0$ (Fig.~\ref{fig4}c), the crossover to a linear dependence at a current threshold $I^*$ (Fig.~\ref{fig4}d),  and the divergence of the quadratic coefficient $b$ for $T \to 0$ (Fig.~\ref{fig4}e). Particularly, the divergence of $b(T \to 0)$ is a strong signature of nodal superconductivity that cannot be easily mimicked by anisotropically gapped superconducting states, providing an additional signature of nodal pairing symmetry~\cite{dahm1996theory,bae2019dielectric, wilcox2022observation}. Similar divergences of $b(T)$ are measured in a second device at optimal dopings of both the electron-like and hole-like superconducting sectors (see Fig.~S8).

Choosing $N=4$ and $\langle|\cos \theta_{I, v_n} |\rangle=0.7$, we can estimate $v_F$ and $v_\Delta$ using Eq.~\ref{eqnl1} and Eq.~\ref{eqnl2}, and \(b(\nu)\) and \(c(\nu)\) extracted from experiments (see Methods for details). The estimated results are shown in Fig.~\ref{fig4}h, where $v_F$ covers a range of $(0.1-0.7) \times 10^5$ m/s, and $v_\Delta$ covers a range of $(0.2-0.5) \times 10^4$ m/s, roughly an order of magnitude smaller than \(v_F\). Remarkably, the density-dependent ratio $v_F/v_\Delta$ estimated here yields similar magnitude ($\simeq 5-20$) and dependence on \(\nu\) to estimates in Fig.~\ref{fig3}e, which was based on an independent analysis of $\rho_s(T)$ measurements. An alternative way to estimate \(v_F\) is to use experimental \(\rho_{s0}(\nu)\) and the aforemetioned expression $\rho_s (\nu)= \frac{1}{8} \sum_{{\bk}} n_k \nabla_{\bk}^2 E = \frac{v_F(\nu) k_F(\nu)}{8 \pi}$. This approach produces a similar but slightly smaller $v_F$ ($\simeq 0.1-0.2 \times 10^5$ m/s), shown as the purple solid line in Fig.~\ref{fig4}h. 

To summarize, our superfluid stiffness measurements provide novel insights into the nature of superconductivity in MATMG. A confluence of signatures, including the linear-in-\(T\) suppression of superfluid stiffness, the observation of the nonlinear Meissner effect, and its temperature dependence provide strong evidence for a nodal pairing symmetry. The linear scaling relation between superfluid stiffness at the zero temperature limit and the superconducting transition temperature suggests an unusual scenario where the superconducting transition is controlled by phase fluctuations rather than Cooper-pair breaking. 

\pagebreak

\section{Methods}

\subsection{Device fabrication}
We prepare the twisted trilayer graphene (TTG) heterostructures using the standard dry-transfer method, where we use stamps consisting of polycarbonate (PC) polymer and polydimethylsiloxane (PDMS) to pick up each
2D material flake sequentially. The TTG heterostructures consist of hBN/graphite/hBN/TTG/hBN/graphite/hBN from top to bottom. Graphite flakes are used as top and bottom gates to control the density and displacement field in TTG. The three pieces of TTG come from the same monolayer graphene, which was pre-cut into three with a laser. 

To make the TTG structure, we pick up each piece sequentially, with a rotation angle \(\theta\) and \(-\theta\) applied to the stage before picking the second and third piece, respectively. After picking up all layers, we drop the stack onto an undoped silicon substrate, which creates a much less parasitic capacitance than a dope one in the rf measurements. We then etch the stack into a Hall bar geometry and deposit Cr/Au contacts through standard nanofabrication lithography. We intentionally make contacts wide, especially for those used for RF measurements, given that RF reflectometry is intrinsically a two-terminal measurement and benefits from wide uniform sample path. We also minimize the area that's between the gold-TTG edge contact region and the double-gated TTG channel region so that the two-terminal signal is dominated by the channel region, in addition to the inevitable contact resistance.

\subsection{Measurements}
Measurements were performed in a Bluefors dilution refrigerator with a base electron temperature of $T=20$~mK, achieved with extensive filtering, shielding, and thermal coupling of the sample within copper packaging. DC transport measurements are performed using the a.c. lock-in technique with an excitation current of 1-10~nA and a low excitation frequency $\sim$ 17~Hz. All DC lines were filtered using a set of RC filters mounted on the 4~K plate and LC filters, to avoid dissipation, mounted at the mixing chamber. The device was embedded in the resonant circuit by way of bias tees and wire bonding the reflectometry line to one of the sample leads, while another lead provided the rf ground. Reflectometry measurements were performed using a vector network analyzer (VNA). The rf excitation signal generated from the VNA was first attenuated by 32 dB via cryostat attenuators installed at various stages of the fridge, and then by another 20 dB as it passed from the coupled-port to the in-port of the directional coupler installed below the mixing chamber. The direction coupler provides a means to remove any spurious resonances in the background of the measurement chain by subtracting the sample signal from measured gain line without the sample. This allows us to precisely quantify the sample resonance shift, and subsequently its superfluid stiffness, without contributions from the measurement chain. After reflection from the LCR circuit, the signal passes through a K\&L filter (450 MHz cutoff), is amplified by 40 dB using a cryogenic amplifier (Weinreb CITLF3) mounted on the 4K plate, and amplified again by 40 dB using a room-temperature amplifier before being measured by the VNA (see Fig. S2). The rf power incident on the sample was varied from -112 dBm to -102 dBm. Measurements were performed with a bandwidth of 0.5-1 kHz, with each VNA trace averaged 5-10 times. 

\subsection{Analytical Model of resonant circuit}
\label{model}
Fig.~S3(a) shows a schematic of the effective microwave resonant circuit, where $L_0$ is the externally added inductor, $C_P$ represents the parasitic capacitance arising primarily from the circuit board, bond wires and the substrate, but may also be added externally in parallel, $R_C$ is the contact resistance and $L_K$ is the inductance of the device in the superconducting state. This circuit can be simplified to an effective series RLC model, shown in Fig.~S3(b) where, using the method of series-parallel transformations, we obtain $R_{eff}=L_0/(R_c C_{eff})$ and $C_{eff}=(C_P - L_0/R_c^2 + L_K/R_c^2)$. Modeling the resonator as a series RLC circuit coupled to a $Z_0=50~\Omega$ transmission line, we can express the impedance of the resonator as:
\begin{equation}
Z= R_{eff} + 2\pi i f L_0 + \frac{1}{2\pi i f C_{eff} } \simeq Q_C Z_0 \left(\frac{1}{Q_I} + i \frac{2 \Delta f}{f_r}\right)
\end{equation}
where the right-hand side is valid near the resonance frequency, $\Delta f = (f-f_r) \ll f_r$. Here $Q_c=Z_{char}/Z_0$ is the coupling quality factor, $Q_I=Z_{char}/R_{eff}$ is the internal quality factor, and the characteristic impedance $Z_{char}=\sqrt{L_0/C_{eff}}$. We also define a loaded quality factor $1/Q_L=(1/Q_C + 1/Q_I)$. 

The reflection coefficient of the resonator, measured by the reflectometry circuit, is given as:
\begin{equation}
\label{eq_Gamma}
\Gamma= \frac{Z-Z_0}{Z+Z_0} = 1 - \frac{2 Q_L/Q_C}{1+2iQ_L (\Delta f/f_r)}
\end{equation}
where the resonance frequency $f_r=1/\sqrt{L_0 C_{eff}} \simeq f_{r0}\left(1+0.5 L_K/(R_c^2 C_P)\right)$, and $f_{r0}=\sqrt{\frac{1}{L_0 C_P} -\frac{1}{R_c^2 C_P^2}}$. In Fig.~S3(c), we numerically evaluate the behavior of the circuit with changing kinetic inductance, for $L_0=205$~nH, $C_P=1.5$~pF, and $R_c=3.5$~k$\Omega$. These values are very typical of a realistic experiment. We compare our analytical expression for $f_r$ with the exact result and obtain a quantitative match for $L_K \leq 500$~nH, the desired operation regime for our experiments, where the frequency shift is linearly related to the kinetic inductance, with $\Delta f_r/f_{r0} = 1/(2 R_c^2 C_P) \Delta L_K$. 
 
\subsection{Circle fitting and analysis}
Reflectometry measurement using a vector network analyzer provides the complex reflection coefficient $S_{21}$ as a function of frequency. Apart from the ideal resonator response described in Eq.~\ref{eq_Gamma}, two additional contributions appear in standard reflectometry measurements~\cite{probst2015efficient}. First, we account for an environmental factor: $a e^{i(\alpha- 2\pi i f \tau)}$, representing an additional
amplitude $a$, a phase shift $\alpha$ and a cable delay $\tau$ determined by the speed of light and the finite length ($\sim$50~mm) of the coaxial cable connecting the device to the direction coupler mounted on the mixing chamber. Second, a phenomenological term $\phi$ is introduced to account for the finite asymmetry of the resonances. This asymmetry is usually small and appears in the complex plane represented by $\Re({S_{21}})-\Im({S_{21}})$ as a rotation of the resonance circle along the real axis by an amount $\phi$~\cite{probst2015efficient}. 

\begin{equation}
\label{eq_circle}
S_{21}= a e^{i(\alpha- 2\pi i f \tau)} \left[1 - \frac{2 (Q_L/Q_C) e^{i\phi}}{1+2 i Q_L (f/f_r-1)}\right]
\end{equation}

We fit our $S_{21}$ data in the complex plane to Eq.~\ref{eq_circle} using the publicly available circle fitting package~\cite{probst2015efficient}. An example of the fitting in the complex plane is shown in Fig. ~S4. The fit parameters $f_r$, $Q_L$, and $Q_C$ are further analyzed using the analytical model described in Sec.~\ref{model}. In Fig.~S5 we show the analysis procedure for converting the resonance frequency into kinetic inductance and superfluid stiffness, as a function of temperature at a representative value of the filling factor $\nu=-2.4$. We can similarly perform a quality factor analysis using $Q_L$ and $Q_C$ and estimate RF resistance ($R_{rf}$) of the sample, using the relation $R_{rf}=\frac{L_0}{Z_0 C_P}(Q_C/Q_L -1)$. At perfect matching, $Q_L=Q_C/2$ such that $R_{rf}=\frac{L_0}{Z_0 C_P}$, the impedance matching criterion we used to design our circuits. In Fig.~S6, we show this analysis as a function of $T$ and $\nu$. $R_{rf} (\nu,T)$ [Fig.~S6(e)] shows good qualitative agreement with both $R_{2t}$ and $R_{4t}$ [Figs.~S6(c,d)]. In the superconducting state, when $R_{4t}=0$, we obtain a quantitative match between $R_{rf}$ and $R_{2t}$ where $R_{rf} \simeq R_{2t} \simeq R_c$ arises purely from the contact resistance $R_c \simeq 3.2$~k$\Omega$.

\subsection{Superfluid stiffness phenomenology and main equations}

Here we derive the formulas for superfluid stiffness of a nodal superconductor, used in the main text. For the temperature and current dependence, we focus on the effects of quasi-particles near the nodes. We note that the effects of classical phase fluctuations can also produce a linear in $T$ suppression of $\rho_s$. However, these are expected to be suppressed below $T_Q\sim \hbar R/L$, which for our parameters ($R\sim 1$ k$\Omega$, $L\sim 100$ nH) yields about $500$ mK \cite{turneaure2000}.

We begin with a mean-field model for superconducting pairing in the active hole bands of TBG. There are two doped hole bands on top of the $\nu = -2$ state. We assume these to be described by a single spin-degenerate band with singlet pairing, for simplicity and concreteness, though the formalism can be adapted to more general scenarios as well. 
\begin{equation}
\begin{gathered}
    \sum_{\bf k} \xi({\bf k}) (c^\dagger_{{\bf k},\uparrow} c_{{\bf k},\uparrow}+c^\dagger_{{\bf k},\downarrow}c_{{\bf k},\downarrow})
    +\Delta({\bf k}) (c^\dagger_{{\bf k}+{\bf Q}/2,\uparrow} c^\dagger_{-{\bf k}+{\bf Q}/2,\downarrow}+c_{-{\bf k}+{\bf Q}/2,\downarrow}c_{{\bf k}+{\bf Q}/2,\uparrow})
    =
    \\
        \sum_{\bf k} 
        \begin{bmatrix}
        c_{{\bf k}+{\bf Q}/2,\uparrow}\\
        c_{-{\bf k}+{\bf Q}/2,\downarrow}^\dagger
        \end{bmatrix}^\dagger
          \begin{bmatrix}
  \xi({\bf k}+{\bf Q}/2) & \Delta({\bf k})\\
  \Delta({\bf k}) & -\xi(-{\bf k}+{\bf Q}/2)
  \end{bmatrix}
          \begin{bmatrix}
        c_{{\bf k}+{\bf Q}/2,\uparrow}\\
        c_{-{\bf k}+{\bf Q}/2,\downarrow}^\dagger
        \end{bmatrix}
  +
  \\
  +
 \sum_{\bf k} \xi(-{\bf k}+{\bf Q}/2),
\end{gathered}
\label{eq:ham0}
\end{equation}
where $\xi({\bf k}) = \varepsilon({\bf k}) -\mu$, where $ \varepsilon({\bf k}) $ - is the single-particle dispersion. The momentum $\bf{Q} = \bf{q}_0 + \bf{q}$ is the momentum of the Cooper pair and $\bf{q}_0$ is the static current of the system.

While the last piece of \eqref{eq:ham0} is a c-number it still contributes to the total energy of the system and needs to be included to avoid convergence issues in the final expressions. Note that, in general, $\Delta(\bk)$ can be dependent on $\bf{Q}$ as well, though we will ignore this dependence for simplicity. This assumption corresponds to neglecting the geometric contribution to the stiffness\cite{peotta2015superfluidity,peotta2023quantum,verma2023geometric}; see e.g. Appendix A2 of Ref. \cite{verma2023geometric} for a discussion in the projected-context. We neglect the geometric stiffness here because in generic models it depends on self-consistent energetics, and thus a microscopic model for the pairing, in addition to the active band wavefunctions \cite{peotta2023quantum}. In the case of the TBG bands we expect the geometric contribution to have the same qualitative dependence on doping as the kinetic contribution, as the dominant kinetic contribution is itself ``geometric": it arises from interactions and the $\bk$ dependence of single-particle wavefunctions. We will discuss this point further in Supplementary materials.  

The superfluid stiffness can be obtained by expanding the grand potential in ${\bf q}$: $[\rho_s]_{\alpha\beta}=\left.\frac{1}{V\hbar^2}\frac{\partial \Omega}{\partial q_\alpha \partial q_\beta}\right|_{{\bf q}=0}$ \cite{peotta2015superfluidity,peotta2023quantum} or calculating the current response. We note two convention differences from prior works: a different definition of Cooper pair momentum, as well as the computation of the superfluid stiffness instead of the superfluid weight $D_s = 4 \rho_s$. These differences correspond to equal and opposite factors of four in the previous expression. 

From Eq. \eqref{eq:ham0} we obtain:
\begin{equation}
    \begin{gathered}
        \Omega \approx -T \sum_{\omega_n, {\bf k}}
        {\rm Tr} \log\Big[i\omega_n -\Delta({\bf k}) \hat{\tau}_1  
        \\
        -\{\xi(\hat{k}_0 ) \hat{\tau}_3+\partial_\alpha\xi(\hat{k}_0) q_\alpha/2+\partial_\alpha\partial_\beta \xi(\hat{k}_0) \hat{\tau}_3 q_\alpha q_\beta/8 \} \Big]
        \\
        +
        \sum_{\bf k}  \xi({\bf k}+{\bf q}_0/2) + \partial_\alpha\xi({\bf k}+{\bf q}_0/2) q_\alpha/2+\partial_\alpha\partial_\beta \xi({\bf k}+{\bf q}_0/2) q_\alpha q_\beta/8,
        \\
        \hat{k}_0= {\bf k}+\hat{\tau}_3 {\bf q}_0/2.
    \end{gathered}
\end{equation}
where $\tau_i$ are matrices in Gor'kov-Nambu space and we assume time reversal symmetry $\xi(-{\bf k}) = \xi({\bf k})$.

The linear in ${\bf q}$ terms vanish after angle integration for ${\bf q}_0=0$ but are nonzero otherwise, corresponding to a finite supercurrent. While $\xi({\bf k})$ vanishes at the Fermi surface, $\partial_\alpha \xi({\bf k})$ and $\partial_\alpha \partial_\beta \xi({\bf k})$ in general do not. This allows us to take ${\bf k}\pm {\bf q}_0/2\approx {\bf k}$ in those derivatives and take $\xi({\bf k}\pm {\bf q}_0/2)\approx \xi({\bf k}) \pm {\bf v}{\bf q}_0/2 $. Then we get:
\begin{equation}
    \begin{gathered}
       [\rho_s]_{\alpha\beta} = [\rho_s^{dia}]_{\alpha\beta}+[\rho_s^{para}]_{\alpha\beta},
       \\
       [\rho_s^{dia}]_{\alpha\beta}
       = \frac{1}{4}
\sum_{{\bf k}}
      \left\{1+
      T\sum_{\omega_n}
      {\rm Tr} [\hat{G}(i\omega_n,{\bf k})\hat{\tau}_3]\right\}
       \partial_\alpha\partial_\beta \xi({\bf k})=
       \\
       = \frac{1}{4} \sum_{{\bf k},\sigma}
    \langle c^\dagger_{\bf k,\sigma}c_{\bf k,\sigma}\rangle\partial_\alpha\partial_\beta \xi({\bf k})
    = \frac{1}{4}
\sum_{\bf k}
\left(
1- \frac{\xi({\bf k})
\tanh\left[\frac{\sqrt{\xi^2({\bf k})+|\Delta({\bf k})|^2}}{2T}\right]
}{\sqrt{\xi^2({\bf k})+|\Delta({\bf k})|^2}}
\right)
\partial_\alpha\partial_\beta \xi({\bf k})
\\
    [\rho_s^{para}]_{\alpha\beta}   = \frac{1}{4} T \sum_{\omega_n, {\bf k}} 
    {\rm Tr}
    [\hat{G}(i\omega_n,{\bf k}) \partial_\alpha \xi  \hat{G}(i\omega_n,{\bf k}) \partial_\beta \xi ],
    \end{gathered}
\end{equation}

where
\begin{equation}
\hat{G}(i\omega_n,{\bf k}) = -\frac{i\omega_n-{\bf v}\cdot {\bf q}_0/2+\xi \hat{\tau}_3+\Delta({\bf k}) \hat{\tau}_1}{[\omega_n+i({\bf v}\cdot {\bf q}_0)/2]^2+\xi^2+\Delta^2({\bf k})}
\end{equation}

For $\partial_\alpha\partial_\beta \xi({\bf k})=const$ (such as for parabolic band), $\rho_s^{dia}$ is simply proportional to the electron density, which is temperature-independent. Ignoring corrections to the chemical potential (assuming $T\ll E_F$), corrections to $\rho_s^{dia}$ can arise from the vicinity of the nodes. However, the leading term will arise from the expansion $\partial_\alpha\partial_\beta \xi({\bf k}) \approx a_0 +a_1 \xi$, so that the integral over momentum will take the form $\int d\xi d\Delta \exp[-\sqrt{\xi^2+\Delta^2}/2T] a_1 \xi^2/\sqrt{\xi^2+\Delta^2}$ and will scale as $\propto T^3$ in temperature.

We thus focus on $\rho_s^{para}$ and its temperature dependence. Evaluating the trace we get:
\begin{equation}
    \begin{gathered}
         [\rho_s^{para}]_{\alpha\beta}   =\frac{T}{2} \sum_{\omega_n, {\bf k}} 
v_\alpha v_\beta
\frac{-[\omega_n+i({\bf v}\cdot {\bf q}_0)/2]^2+\xi^2+\Delta^2({\bf k})}{([\omega_n+i({\bf v}\cdot {\bf q}_0)/2]^2+\xi^2+\Delta^2({\bf k}))^2}
\\
=
-\frac{1}{16T} \sum_{\bf k} v_\alpha v_\beta \left( \frac{1}{\cosh^2\frac{\sqrt{\xi^2+\Delta^2({\bf k})}+({\bf v}\cdot {\bf q}_0)/2}{2T}}
+
\frac{1}{\cosh^2\frac{\sqrt{\xi^2+\Delta^2({\bf k})}-({\bf v}\cdot {\bf q}_0)/2}{2T}}\right).
    \end{gathered}
\end{equation}
Since the contribution is exponentially suppressed by the gap, we focus on the vicinity of the nodes, where $\Delta({\bf k})\approx {\bf v}_\Delta \cdot ({\bf k}-{\bf k}_F)$ and $\xi\approx{\bf v} \cdot ({\bf k}-{\bf k}_F)$.

\begin{equation}
\begin{gathered}
\delta \rho_s(T,I)/\text{1 node} / =
[\rho_s^{para}]_{\alpha\beta} \approx
\\
\approx
\sum_n
-\frac{v_n^\alpha v_n^\beta/ (v_n v_\Delta)}{16T \sin \theta_{v,v_\Delta}} \int \frac{d\xi d\Delta}{(2\pi)^2}\left( \frac{1}{\cosh^2\frac{\sqrt{\xi^2+\Delta^2}+({\bf v}_n\cdot {\bf q}_0)/2}{2T}}
+
\frac{1}{\cosh^2\frac{\sqrt{\xi^2+\Delta^2({\bf k})}-({\bf v}_n\cdot {\bf q}_0)/2}{2T}}\right)
=
\\
=
-
\sum_n
\frac{v_n^\alpha v_n^\beta/ (v_n v_\Delta)}{4\pi \sin\theta_{v,v_\Delta}} T
\log
\left[
2 \cosh \frac{({\bf v}_n\cdot {\bf q}_0)}{4T}
\right],
\end{gathered}
\label{eq:gen_res}
\end{equation}
where ${\bf q}_0\propto{\bf I}$ (see Eq. S2). Thus, at zero current, one gets $\delta \rho_s\propto -T$, while at $({\bf v}\cdot {\bf q}_0)\gg T$ one gets $\delta \rho_s \propto -I$. Note that expression \ref{eq:gen_res} can be modified by Fermi-liquid effects; these result in an additional multiplicative factor in the above formula \cite{MILLIS19981742}; in cuprates the value of this factor was found to be close to 1 \cite{ioffe2002d,wenrmp}.

The results simplify considerably for: (a) symmetry-protected nodes, i.e. ${\bf v} \perp {\bf v}_\Delta$ (b) non-nematic SC state, i.e. $\rho_{\alpha\beta} \propto \delta_{\alpha\beta}$. From (b) it follows that $\sum_n (v_n^\alpha)^2 = N v^2/2$, where $N$ is the number of nodes. For $N=4$ (d-wave) we recover $\delta \rho_s (T,q_0=0)  = -\frac{ v \log 2}{2\pi v_\Delta} T$\cite{lee1997unusual}.

Let us now estimate how the current-induced correction depends on temperature. We use $q_0 = \frac{I}{2e  \rho_s^0W}$ from Eq. S2 and expand \eqref{eq:gen_res} using the assumptions (a) and (b). We further compute the $\rho_s\approx {\rm Tr} \rho_{\alpha\beta}/2$ to average over current-induced anisotropies. The result is
\begin{equation}
\begin{gathered}
    \delta \rho_s (T\gg v q_0)
    =
    -\frac{N v }{8\pi v_\Delta} T
    \left(\log 2
    +
    \frac{v^2 I^2}{16 (4e \rho_s^0W)^2 T^2}
    \right)  
    \end{gathered}
    \label{eq:lint}
\end{equation}
Now let's express $\delta \rho_s$ at large current (compared to the effects of  temperature):
\begin{equation}
    \delta \rho_s (v q_0\gg T)= \frac{N v^2}{64 v_\Delta \pi}
    \frac{I}{e \rho_s^0 W}
    \langle\cos \theta_{I, v_n}\rangle
     \label{eq:lincur}
\end{equation}

\pagebreak

\backmatter

\bmhead{Acknowledgements}
We thank Steve Kivelson, Erez Berg, Amir Yacoby, and Marie Wesson for helpful discussions. The major experimental work is supported by ARO MURI (W911NF-21-2-0147). P.K. and A.B. acknowledge support from the DOE (DE-SC0012260). AV, PL and PAV are supported by a Simons Investigator grant (AV) and by NSF-DMR 2220703. PAV acknowledges support by a Quantum-CT Quantum Regional Partnership Investments (QRPI) Award.  K.W. and T.T. acknowledge support from the JSPS KAKENHI (Grant Numbers 21H05233 and 23H02052) and World Premier International Research Center Initiative (WPI), MEXT, Japan. Nanofabrication was performed at the Center for Nanoscale Systems at Harvard, supported in part by an NSF NNIN award ECS- 00335765.

\clearpage

\clearpage 

\bibliography{sn-bibliography}

\pagebreak

\setcounter{equation}{0}
\renewcommand{\theequation}{S\arabic{equation}}
\setcounter{figure}{0}
\renewcommand{\thefigure}{S\arabic{figure}}

\section{Superfluid stiffness of twisted multilayer graphene superconductors: Supplementary Material}

\subsection{Device DC transport characterization}

We characterize the TTG device presented in the main text through 4-terminal DC transport measurements, which show results consistent with our previous study on magic-angle TTG. Fig.~S1a shows the 4-terminal resistance \(R\) as a function of \(\nu\) and
magnetic field \(B\) at zero displacement field and a temperature of 2 K. Characteristic of magic-angle TTG, there appear two sets of Landau fans. One set has fans emerging from integer fillings of the TTG flat bands. The sequences in these fans generally have larger slopes in \(B-\nu\) plane, in contrast to the other set that appears at low magnetic fields as "arc-like" quantum oscillations features. These features emerge from the dispersive Dirac cone sector of the TTG electron bands, in addition to the flat band sector. 

These two distinct quantum oscillation features with large and small slopes, respectively, are because the flat and Dirac cone sectors are filled simultaneously and a larger proportion of carriers goes into the flat bands until they are fully filled. We can calculate the moiré unit cell area \(A_m\) based on the flat band fan sequences, which satisfy \(BA_m/\phi_0=C\nu+s\), where \(\phi_0=e/h\) is the magnetic flux quantum, \(C\) is the Chern number (and the slope), and \(s\) is the intercept on the \(\nu\) axis from which the fans emerge. Then we can obtain the twist angle \(\theta=1.55\)\textdegree~from the relation \(A_m=\frac{\sqrt{3}a^2}{2\theta}\), where \(a=0.246\)~nm is the graphene lattice constant. We have measured the fan diagrams for multiple pairs of contacts in different areas of the sample and they all show similar features and twist angles, indicating the TTG sample is highly uniform. Fig.~S1b shows \(R\) as a function of \(\nu\) and displacement field \(D\) at a dilution fridge base temperature of 30~mK, with a resistive feature at the charge neutrality and superconducting states within \(-3<\nu<-2\) on the hole-doped side and \(1.5<\nu<3\) on the electron-doped side.

\subsection{Superfluid Stiffness, superfluid Density and kinetic Inductance}

Here we write out the general relations between superfluid stiffness $\rho_s$, superfluid weight $D_s$ and kinetic inductance. In particular, the change in free energy of a superconductor under current is related to Cooper pair momentum as \cite{wenrmp,peotta2015superfluidity}:

\begin{equation}
    \Delta F = S D_s \frac{Q^2}{8} 
    =S \rho_s \frac{Q^2}{2}
    \equiv \frac{L_{kin} I^2}{2},
\end{equation}
where $S$ is the area of the sample and $Q$ is the Cooper pair momentum [note that definition of superfluid weight \cite{ioffe2002d} is 4 times smaller than of superfluid density $\rho_s = D_s/4$]. We will assume a rectangular sample with $S=l W$, $l$ being the length (along the current direction) and $W$ - width (perpendicular to the current).

Current can be obtained from gauge coupling $Q \to Q - \frac{2 e}{c} \vec{A}$ and using $j_{2D} = - c \frac{\partial \Delta F}{ \partial A}$ \cite{wen1997}:

\begin{equation}
    I = j_{2D} W =  2e \rho_s Q,
    \label{eq:current-momentum}
\end{equation}

which leads to the relation:
\begin{equation}
  L_{kin} = \frac{l}{4 e^2 \rho_s W}.
\end{equation}

For the case of a Gallilean invariant superconductor where $m_{CP} = 2 m_e$ one gets $D_s = n_e/m_e$, $\rho_s = n_e/(4 m_e)$ and $L_{kin} = \frac{l m_e}{n_e e^2 n W}$.

\subsection{Dispersion on top of correlated insulator}

In this section we describe how we calculate the hole dispersion on top of the $\nu = -2$ correlated insulator. While we will make several assumptions in order to arrive at a simple, easy to compute, dispersion, we will later comment on its essential features and why we expect them to be generic amongst most $\nu = -2$ states.

We will take the ``strong coupling" limit\cite{bultinckGroundStateHidden2020,vafekRenormalizationGroupStudy2020,lianTBGIVExact2020,ledwithStrongCouplingTheory2021}, under which the interaction strength is much larger than the bare single-particle dispersion but much smaller than the band gap to the remote bands. This is a good approximation when the bands are flat: close to the magic angle and for not-too-large sample strain. The resulting dispersion we obtain will be entirely generated from interactions and the $\bk$-dependence of the single particle wavefunctions (``quantum geometry"). It is convenient to work in the sublattice or Chern basis in this limit, which consists of four bands with $C=+1$ and four bands with $C=-1$, each set of which can be labeled by spin and valley. The Chern basis is convenient for Coulomb interactions because the form factors, which determine the projected density operators
\begin{equation}
    \Lambda_{\bq \alpha \beta}(\bk) = \langle u_{\bk \alpha}u_{\bk + \bq \beta}\rangle \quad \rho_\bq = \sum_\bk c^\dag_{\bk} \Lambda_{\bq}(\bk) c_{\bk + \bq},
\end{equation}
are approximately diagonal,
\begin{equation}
    \Lambda_{\bq}(\bk) = \begin{pmatrix} \lambda_{\bq}(\bk) I_{4 \times 4} & 0 \\ 0 & \overline{\lambda_{\bq}(\bk)} \end{pmatrix}.
\end{equation}

The Hamiltonian is 
\begin{equation}
    \H = \frac{1}{2A} \sum_\bq V_\bq \delta \rho_{\bq} \delta \rho_{-\bq}, \qquad \delta \rho_\bq = \rho_\bq - \rho_\bq^{\rm CNP},
\end{equation}

where $\rho_\bq^{\rm CNP} = \frac{1}{2}\sum_{\bG} \delta_{\bq, \bG} \mathrm{tr}\, \Lambda_\bq(\bk)$ is half the total density of the flat bands and $A$ is the sample area. There is a $U(4) \times U(4)$ symmetry of $\H$ consisting of rotating the bands in each Chern sector into each other, which is broken by off-diagonal terms in the form factors, single particle dispersion, strain, and other corrections such as those from $K$-point phonons. 

Exact eigenstates of $\H$ are generalized quantum Hall ferromagnets, consisting of filling entirely some number of the eight Chern bands. This large manifold of states is split by the aforementioned symmetry breaking perturbations, and the resulting ground state is determined by their competition. In this supplement we will be agnostic about the precise $\nu = -2$ ground state. Thus we will simply use the symmetric hole dispersion, which is the same for all strong coupling ground states in the $U(4) \times U(4)$ symmetric limit and can be computed exactly. The dispersion $h_\nu$ for holes at filling $\nu$ is then
\begin{equation}
\begin{aligned}
h_{\nu}(\bk) & = -h_F(\bk) + \nu h_H(\bk), \\
    h_F(\bk) & = \frac{1}{2A} \sum_{\bq} V_\bq |\lambda_\bq(\bk)|^2, \\
    h_H(\bk) & =  \frac{1}{A} \sum_{\bG} V_\bG \Lambda_\bG(\bk) \sum_{\bk'} \lambda_{-\bG}(\bk'),
    \end{aligned}
\end{equation}
where $h_H$ and $h_F$ are the Hartree and Fock contributions respectively. 

Both $h_H$ and $h_F$ are strongly peaked in a small region near $\bk = \Gamma$, and quite flat away from $\Gamma$. We expect this to be a generic feature of most band dispersions of translationally symmetric correlated insulators. Indeed, away from $\Gamma$, the wavefunctions are largely $\bk$ independent and are well described by linear combinations of Wannier states localized at the AA sites of the moiré lattice. The region near the Gamma point is distinct, since $\psi_\Gamma(\br = AA)$ vanishes, and has strong $\bk$ dependence \cite{guineaElectrostaticEffectsBand2018,carrDerivationWannierOrbitals2019,rademakerChargeSmootheningBand2019,ceaBandStructureInsulating2020,goodwinHartreeTheoryCalculations2020,kangCascadesLightHeavy2021,songMagicAngleTwistedBilayer2022,pierceUnconventionalSequenceCorrelated2021,parkerFieldtunedZerofieldFractional2021}. Thus, any translationally-symmetric mean-field potential projected to the flat bands, $ \Delta(\bk) = \langle u_{\bk}| \hat{\Delta} |u_\bk\rangle$, will in general be flat away from $\Gamma$ and have a bump or dip near $\Gamma$. The effect is especially intuitive for the Hartree dispersion, where $\Gamma$-point electrons are cheaper to dope than the rest of the Brillouin Zone because their wavefunctions vanish at the AA charge density peak, thereby evading a large electrostatic penalty. We pause to comment that the intervalley kekulé spiral state is a translation-breaking candidate state stabilized by strain that may have a distinct dispersion\cite{kwanKekulSpiralOrder2021,wagnerGlobalPhaseDiagram2022,wangGroundstateOrderMagicangle2023,nuckollsQuantumTexturesManybody2023a,kimImagingIntervalleyCoherent2023}. As we alluded to in the previous section, we expect that the geometric contribution to the superfluid stiffness will have a similar doping dependence as the interaction-induced kinetic part. Indeed, both the dispersion of the band and the quantum geometry are generated by the strong $\bk$-dependence of the wavefunctions near the $\Gamma$ point band edge.

In our computations of the Hartree Fock dispersion we use gate screened Coulomb interactions $V_\bq = \frac{1}{2\epsilon \epsilon_0 q} \tanh qd$ with gate distance $d = 20$nm, a relative permittivity $\epsilon = 20$ (coming from screening of hBN, remote band electrons, and the decoupled Dirac cone sector). Our BM model wavefunctions correspond to a twist angle of $1.5^\circ$ and AA-AB tunneling ratio of $w_0/w_1 = 0.5$. 

We see that the resulting band structure is highly non-parabolic. To visualize the non-parabolicity, we plot $\frac{1}{2} v_F k_F$ and $E_F$, equal for a parabolic band, versus the filling $\nu$. See Fig.\ref{fig:nonparab}. While the Fermi energy monotonically increases, $\frac{1}{2} v_F k_F$ peaks at a small filling factor and subsequently decreases. Note that the kinetic contribution to the superfluid stiffness is $\frac{1}{8\pi} v_F k_F$.

\subsection{Superfluid stiffness in other devices}
We measured a total of 4 devices comprising three twisted trilayer graphene devices (TTG1, TTG3, TTG-MIT) and one twisted quadrilayer graphene device (T4G). Within these devices, six independent superconducting domes, including both electron (e) and hole (h) doped sectors were investigated using simultaneous DC transport and rf-reflectometry. In Fig. S7, we show the zero-temperature superfluid stiffness $\rho_s(0)$ as a function of $T_C$ for all measured superconducting domes. A roughly linear trend is observed, even for devices where twist-disorder appears to be strong (see discussion below). 

In Figs. S8 and S9 we show data from all measured superconducting domes where the top panel represents resistance measurements as a function of $\nu$ and $T$, the middle panel is the frequency shift of the resonator and the bottom panel is the temperature derivative of stiffness ($d\rho_s/dT$) or equivalently the frequency shift ($df_r/dT$). The derivative plots allow us to identify regions within the superconducting dome where the stiffness varies strongly as a function of temperature, with red regions indicating zero slope. All measured superconducting domes show large values of $d\rho_s/dT$ at low temperatures, consistent with nodal superconductivity. However, in some devices, for certain values of $\nu$, we measure $d\rho_s/dT \simeq 0$ representing a saturated $\rho_s(T)$ at low temperature (bottom panel of Figs. S7(b) and S7(c)). On inspection, we find that these regions coincide with fillings where the two-terminal resistance also shows signatures of large twist-angle disorder, identified as a cascade of resistance transitions as a function of temperature and doping (top panel of Figs. S7(b) and S7(c)). For example, in T4G-e, we observe saturated $\rho(T)$ in the same region of $(\nu,T) \simeq (2-3, 0.02-0.4 K)$ where $R_{2t}$ shows signatures of strong twist-disorder induced inhomogeneity in DC transport. Similar behavior is seen in TTG-MIT where we observe T-linear suppression of $\rho_s(T)$ in the gate range $V_{G}=17-19$~V where $T_c 
\simeq 0.7$~K is highest, but a low-temperature saturation of $\rho_s(T)$ is observed in the gate range $V_{G}=13-17$~V, where $T_c\simeq 0.3$~K is lower, and also cascades of resistance transitions are observed in $R_{2t}$. Our data suggests that strong twist disorder tends to produce a low-temperature saturation of $\rho_s(T)$ whereas regions of clean superconductivity show linear-T behavior consistent with nodal superconductivity. 

In Fig. S10 we show the nonlinear Meissner effect observed at optimal doping in TTG1 in both the electron and hole doped sectors. Red dashed lines show fits to a quadratic dependence, $\delta \rho_s=-bI_b^2$. The quadratic coefficient $b$ as a function of $T$ shows diverging behavior as $T \to 0$, indicating nodal superconductivity. Nonlinear Meissner effect measurements are particularly sensitive to the magnitude of the critical current and require relatively large values of $I_c \geq$100~nA. Moreover, twist-disorder-induced inhomogeneity causes cascades of superconducting transitions as a function of $I$ that are not conducive to a reliable investigation of $\rho_s(I)$. Only the cleanest devices, TTG1 and TTG3, showed the nonlinear Meissner effect, whereas in the more twist-disordered devices we were not able to perform reliable $\rho_s(I)$ measurements.

\clearpage

\begin{figure}[H]
\includegraphics{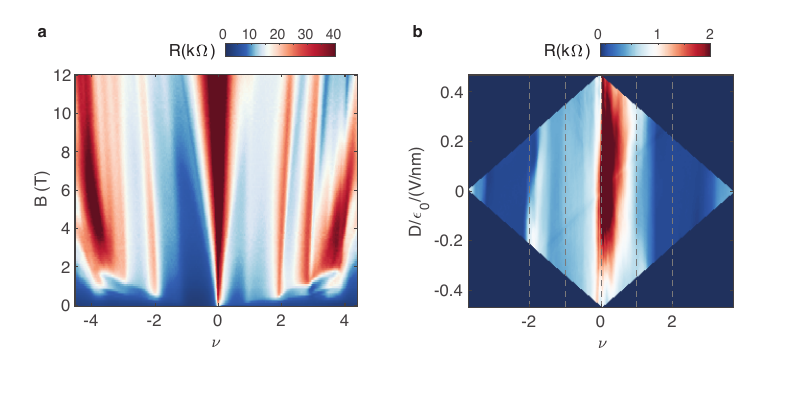}
\label{fig_DC_transport}
\caption{{\bf Device DC transport characterization.} \textbf{a,} TTG 4-terminal \(R\) as a function of \(\nu\) and magnetic field \(B\) at zero displacement field and a temperature of 2~K. The Landau fans show two set of structures: one set with large slopes and the other that appear at low \(B\) with shallow slopes. \textbf{b,} TTG 4-terminal \(R\) as a function of \(\nu\) and displacement field \(D\) at zero \(B\) and a temperature of 30~mK. }
\label{fig_DC_transport_caption}
\end{figure}

\begin{figure}[H]
\includegraphics[width=1\textwidth]{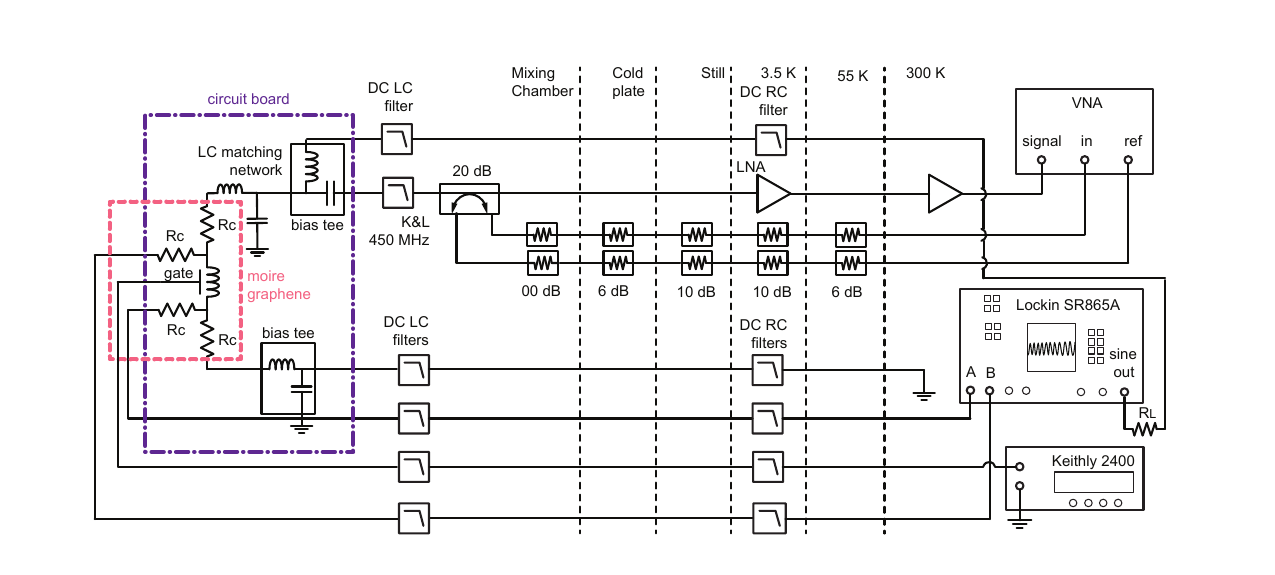}
\label{fig_matrix}
\caption{{\bf Measurement setup} Circuit model of measurement setup for superfluid stiffness measurements. The moiré graphene sample is indicated by dashed pink lines while the circuit board--containing sample, LC matching network, and bias tees to allow impedance matching for microwave measurement--is indicated by dashed purple lines. DC and RF filtering is used to reduce noise throughout the measurement chain. The RF signal is sent through an attenuated input line before entering a directional coupler. Sample contact resistance of  $\simeq 3.2$~k$\Omega$ is given by $R_c$. Graphite top and bottom gates are represented by "gate". The final signal is amplified both at 4~K and room temperature before being measured by a vector network analyzer (VNA).}
\label{fig_fr}
\end{figure}

\begin{figure}[H]
\centering
\includegraphics[width=1\textwidth]{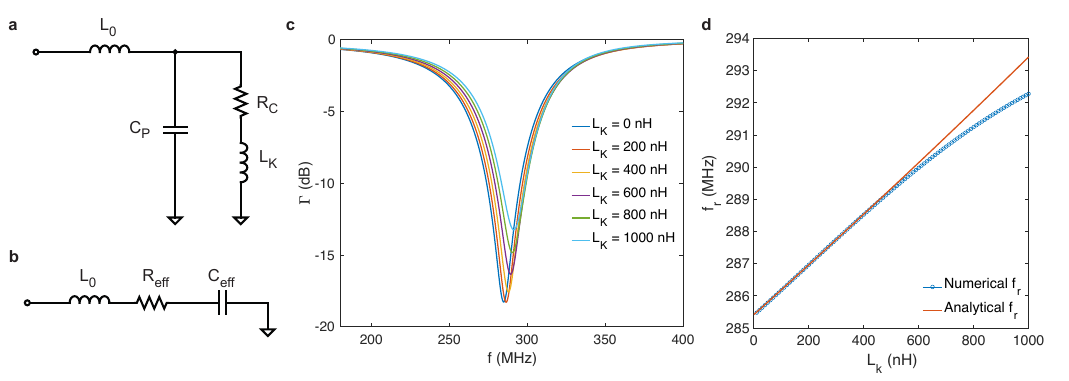}
\caption{{\bf Analytical circuit model:} (a) Model of the resonant circuit with an externally added inductor $L_0$, capacitance $C_P$, contact resistance $R_C$ and the kinetic inductance $L_K$ (b) Series RLC circuit equivalent to the circuit in (a). (c) Numerical simulation of the circuit in (a) for realistic parameters and changing kinetic inductance. (d) Resonance frequency $f_r$ comparing the linear analytic formula presented above and the exact result.}\label{fig_am}
\end{figure}

\begin{figure}[H]
\includegraphics[width=1\textwidth]{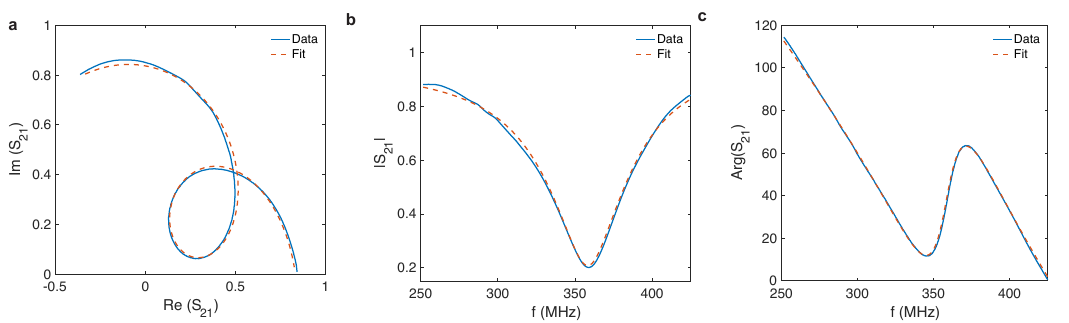}
\label{fig_circle}
\caption{{\bf Circle fitting:} (a) Circle fitting of  $S_{21}$ in the complex plane. (b) Fitting of the reflection amplitude $|S_{21}(f)|$. (c) Fitting of the phase expressed in degrees $\arg{S_{21}(f)}$. }
\label{fig_fr}
\end{figure}

\begin{figure}[H]
\includegraphics[width=1\textwidth]{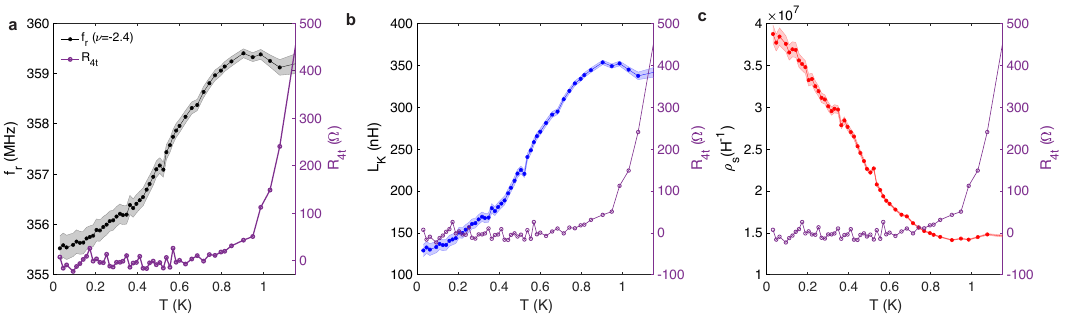}
\label{fig_freq_shift}
\caption{{\bf Frequency shift analysis:} (a) Resonance frequency $f_r$ as a function of $T$ measured at $\nu=-2.4$ compared with simultaneously measured 4-terminal DC resistance $R_{4t}$. The strong downward shift of $f_r$ at $T \simeq 0.9$~K coincides with the onset of zero four-terminal resistance. (b) $L_K$ extracted from $f_r$ shows large kinetic inductances $\simeq 120-350$~nH. (c) Superfluid stiffness extracted from $L_K$, accounting for the geometric aspect ratio of the device $\rho_s=(w/l)1/L_K$, where $w/l \simeq 5$.}\label{fig_fr}
\end{figure}

\begin{figure}[H]
\includegraphics[width=1\textwidth]{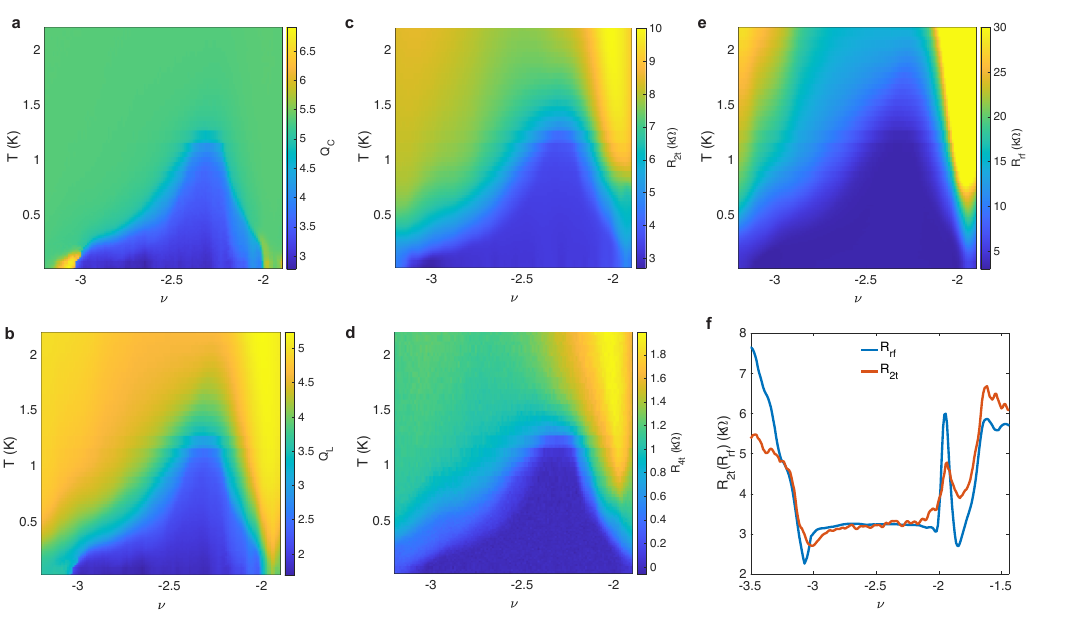}
\label{fig_qfactor}
\caption{{\bf Quality factor analysis:} (a) Coupling quality factor $Q_C$ and (b) loaded quality factor $Q_L$ obtained from circle fitting. (c) Simulatanously measured two-terminal resistance $R_{2t}$ and (d) four-termianl resistance $R_{4t}$. (e) Extracted rf resistance $R_{rf}$ showing a qualitative agreement with (c) and (d) in the $\nu-T$ plane. (f) Quantitative agreement is achieved in the superconducting state where $R_{rf} \simeq R_{2t} \simeq R_c$, where $R_c$ is the contact resistance.}
\label{fig_fr}
\end{figure}

\begin{figure}[H]
\includegraphics[width=1\textwidth]{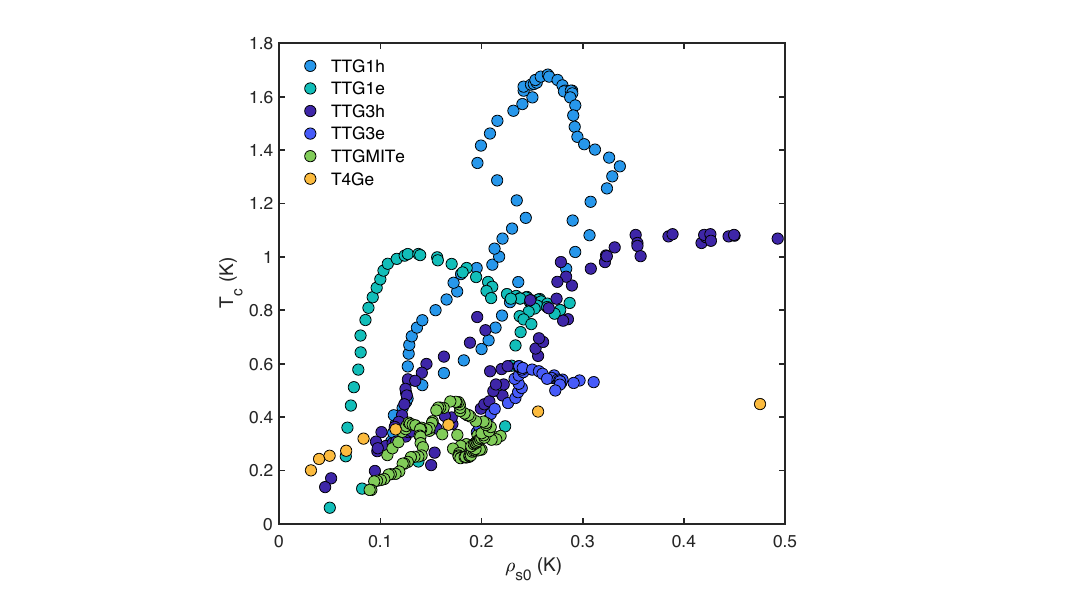}
\label{fig_Uemura}
\caption{{\bf Uemura's law:} Scaling of $\rho_{s0}$ versus $T_c$ (obtained from resistance measurements) for both electron and hole side superconductors in all measured devices. A roughly linear trend is generally observed, even for devices with twist-angle disorder.}
\label{fig_fr}
\end{figure}

\begin{figure}[H]
\includegraphics[width=1\textwidth]{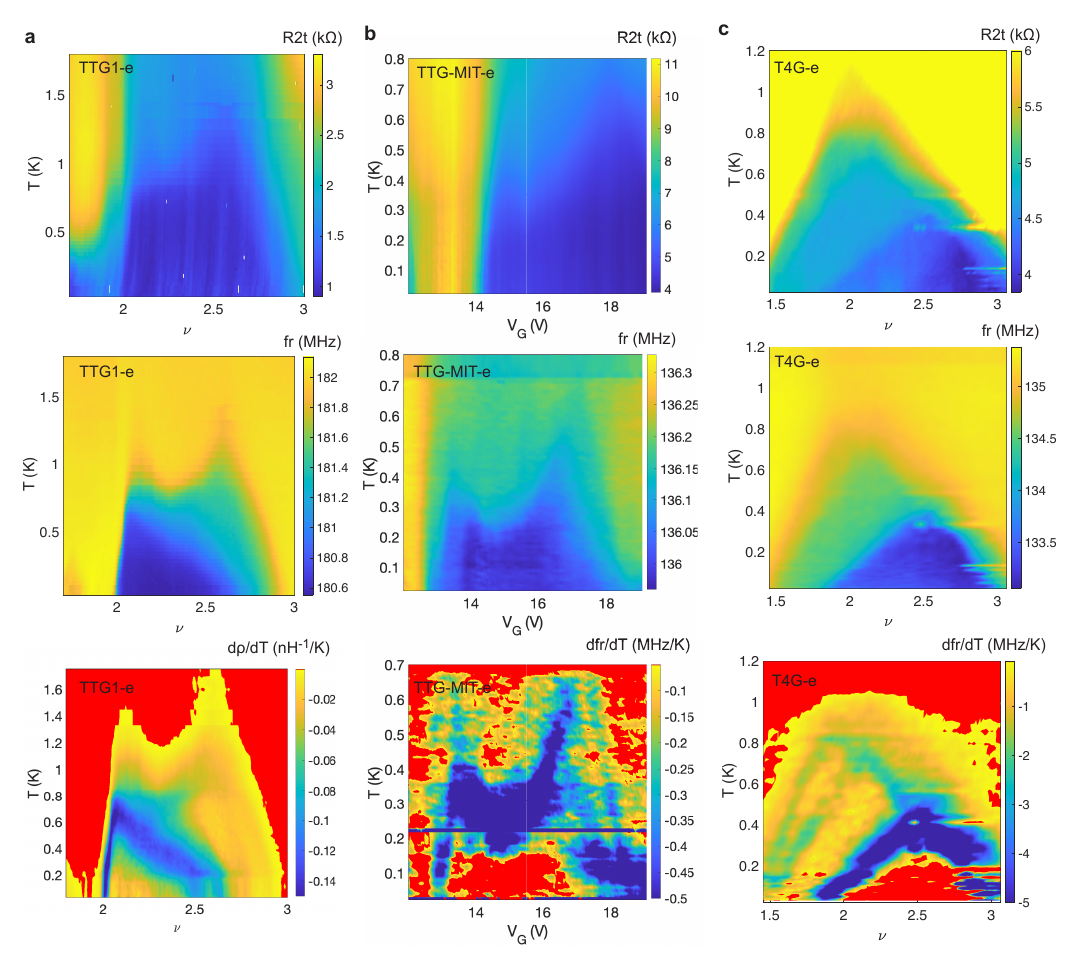}
\label{fig_matrix}
\caption{{\bf Superconducting domes in other devices:} Superconducting domes measured as a function of temperature $T$ and carrier density for (a) TTG1-e (b)TTG-MIT-e and (c) T4G-e. Top panel: Two terminal resistance $R_{2t}$ measured with DC transport. Middle panel: Resonance frequency $f_r$ measured simultaneously with rf-reflectometry. Bottom panel: Temperature derivative of the superfluid stiffness, $d\rho_s/dT$ (for TTG1-e), or equivalently the resonance frequency, $df_{r}/dT$ (for TTG-MIT-e and T4G-e). $d\rho_s/dT=0$ ($df_{r}/dT =0$) is indicated with red on the colorscale. The smaller frequency shift in the latter devices makes $df_{r}/dT$ more reliable than $d\rho/dT$. We observe non-zero derivatives within most of the $\nu-T$ phase diagram, especially at optimal doping (max Tc) for every superconducting dome. At optimal doping, all superconducting domes show linear-T low-temperature stiffness variation, with a roughly constant slope. On the other hand, away from optimal doping, several parts of the $\nu-T$ phase diagram, especially in TTG-MIT-e and T4G-e show strong resistivity variations within the superconducting dome, identified from the two-terminal resistance measurements (top panel). We attribute these resistivity variations to twist angle disorder, where multiple superconducting grains with different transition temperatures produce smaller overlapping superconducting domes. Corresponding twist-disordered parts of the phase-diagram usually show saturated frequency dependence as a function of temperature, indicated in the bottom panel as regions with zero slope (red on the colorscale).}
\label{fig_fr}
\end{figure}

\begin{figure}[H]
\includegraphics[width=1\textwidth]{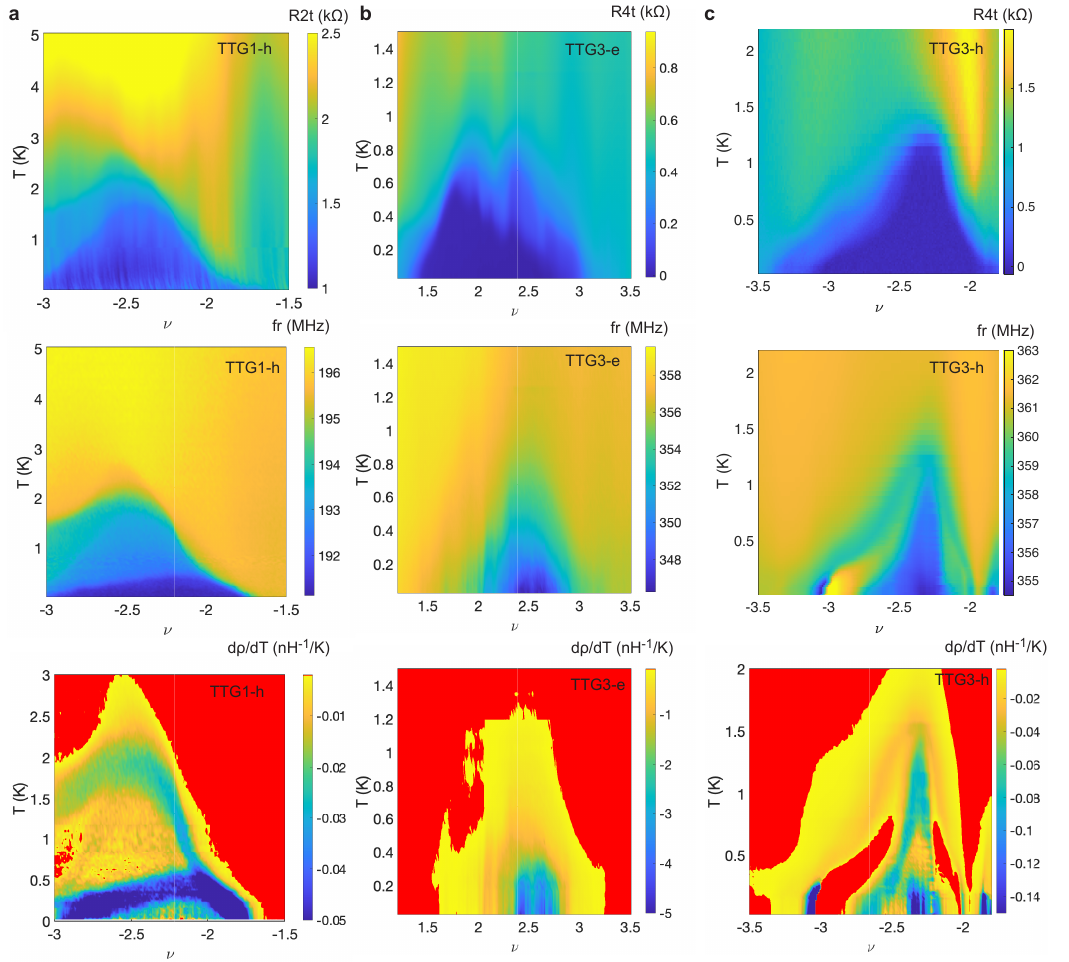}
\label{fig_matrix}
\caption{{\bf Superconducting domes in other devices:} Superconducting domes measured as a function of temperature $T$ and carrier density for (a) TTG1-h (b)TTG3-e and (c) TTG3-h (discussed in the main text). Top panel: Device resistance ($R_{2t}$ or $R_{4t}$) measured with DC transport. Middle panel: Resonance frequency $f_r$ measured simultaneously with rf-reflectometry. Bottom panel: Temperature derivative of the superfluid stiffness, $d\rho_s/dT$, with $d\rho_s/dT=0$ indicated with red on the colorscale. We observe non-zero derivatives within most of the $\nu-T$ phase diagram, especially at optimal doping (max Tc) for every superconducting dome.}
\label{fig_fr}
\end{figure}

\begin{figure}[H]
\includegraphics[width=1\textwidth]{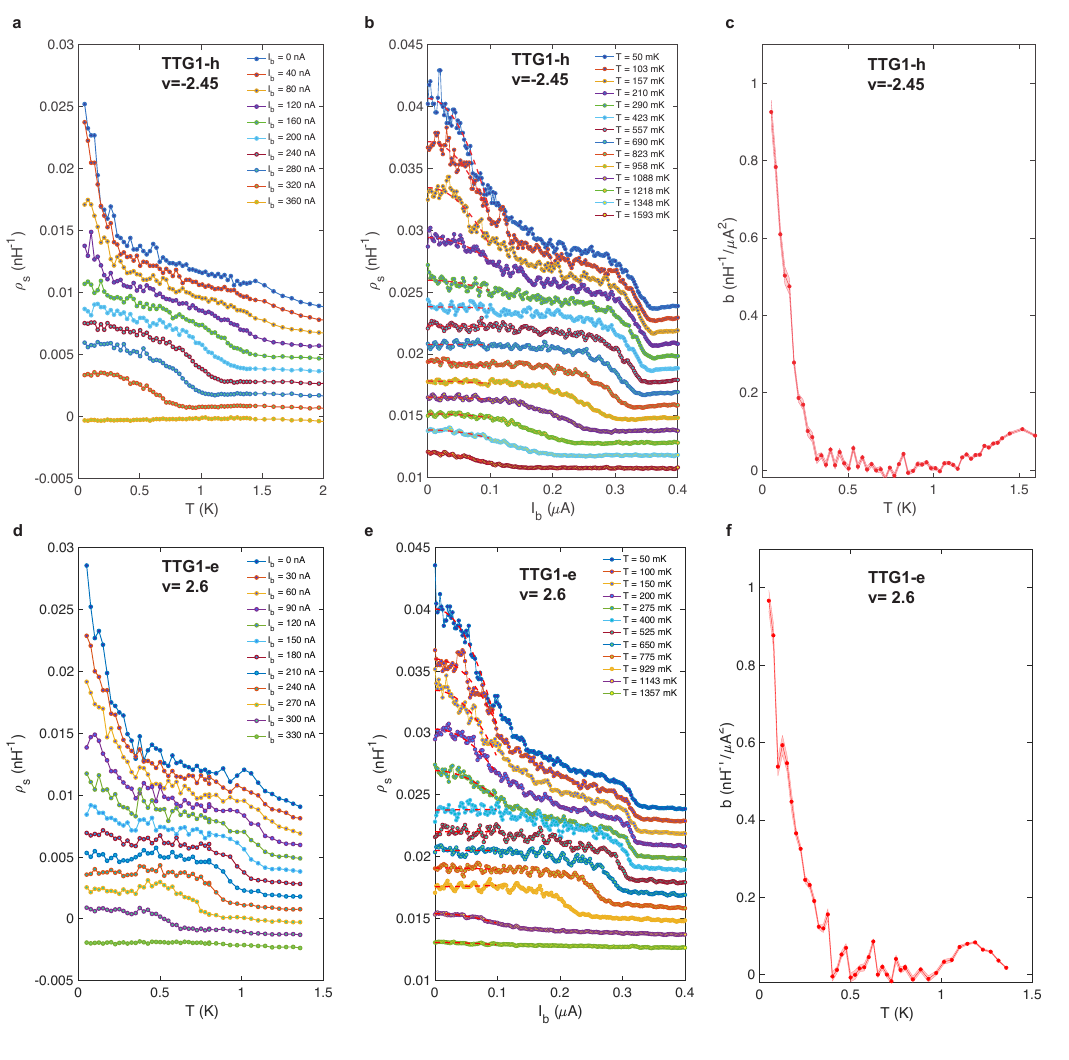}
\label{fig_matrix}
\caption{{\bf Nonlinear Meissner effect at optimal doping in TTG-1:} (a-c) Nonlinear Meissner effect measurement at optimal hole doping in TTG-1, $\nu=-2.45$ (a) $\rho_s(T)$ measured at different values of supercurrent bias $I_b$. (b) $\rho_s(I)$ measured at different values of temperature $T$. Red dashed lines show fits to a quadratic dependence, $\delta \rho_s=-bI_b^2$ (c) Quadratic coefficient $b$ as a function of $T$, show diverging behavior as $T \to 0$. (d-f) Nonlinear Meissner effect measurement at optimal electron doping in TTG-1, $\nu=+2.6$ (a) $\rho_s(T)$ measured at different values of supercurrent bias $I_b$. (b) $\rho_s(I)$ measured at different values of temperature $T$. Red dashed lines show fits to a quadratic dependence, $\delta\rho_s=-bI_b^2$ (c) Quadratic coefficient $b$ as a function of $T$, show diverging behavior as $T \to 0$.}
\label{fig_fr}
\end{figure}

\begin{figure}[H]
    \centering
    \includegraphics[width=0.7\textwidth]{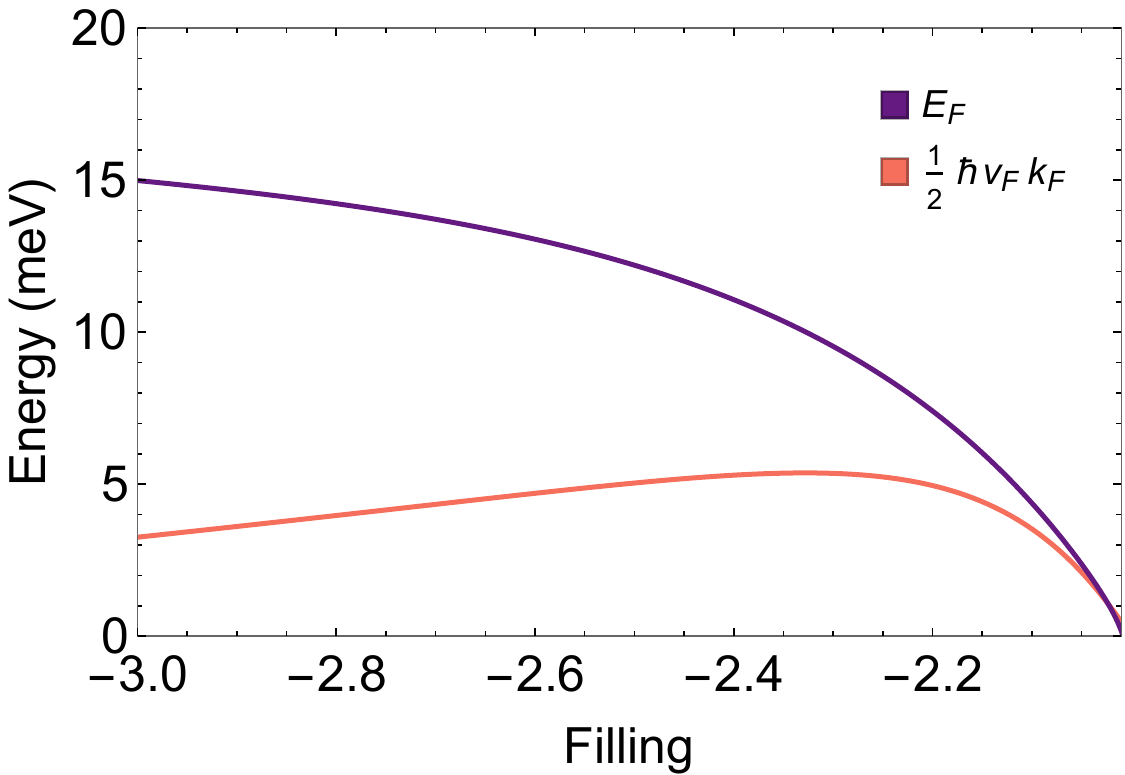}
    \caption{{\bf{Non-parabolicity of Hartree Fock band structure}}. While $E_F = \frac{1}{2} \hbar v_F k_F$ for a parabolic band, the Hartree Fock dispersion of the TTG flat bands is much more concentrated at the $\Gamma$ point, and flattens out elsewhere. Thus, $v_F$ begins decreasing at rather-small doping. For a non-parabolic band, $E_F \neq \frac{1}{2} \hbar v_F k_F = 4\pi \rho_{s0}^{dia}$.}
    \label{fig:nonparab}
\end{figure}

\end{document}